\documentclass[useAMS,usenatbib]{mn2e}
\usepackage{pifont}
\usepackage{graphicx}
\usepackage[hyphens]{url}
\usepackage{multirow}
\usepackage{mathtools}
\usepackage{textcomp}
\usepackage{amssymb}
\usepackage{amsmath}
\usepackage{color}
\usepackage{abbrevs}
\usepackage{xspace}
\usepackage{float}
\usepackage[usenames,dvipsnames]{xcolor}
\begingroup
  \catcode`\_=\active
  \gdef_#1{\ensuremath{\sb{\mathrm{#1}}}}
\endgroup
\mathcode`\_=\string"8000
\catcode`\_=12

\newcommand{\rev}{\textcolor{Black}}

\newcommand{\Msolar}{M$_{\odot}$\xspace}

\newcommand{\Rsolar}{R$_{\odot\xspace}$\xspace}

\newcommand{\atcc}{atoms/cm$^{3}$\xspace}

\newcommand{\simname}{\textsc}
\newcommand{\Stromgren}{Str\"{o}mgren\xspace}

\newcommand{\tick}{$\checkmark$}

\newabbrev\ISM{Interstellar Medium (ISM)}[ISM]
\newabbrev\CSM{Circumstellar Medium (CSM)}[CSM]
\newabbrev\WNM{Warm Neutral Medium (WNM)}[WNM]
\newabbrev\WIM{Warm Ionised Medium (WIM)}[WIM]
\newabbrev\CNM{Cold Neutral Medium (CNM)}[CNM]
\newabbrev\IMF{Initial Mass Function (IMF)}[IMF]
\newabbrev\AMR{Adaptive Mesh Refinement (AMR)}[AMR]
\newabbrev\HGB{Horizontal Giant Branch (HGB)}[HGB]

\makeatletter
\renewcommand\maybe@space@{%
  \maybe@ictrue % <= this is new
  \expandafter   \@tfor
    \expandafter \reserved@a
    \expandafter :%
    \expandafter =%
                 \nospacelist
                 \do \t@st@ic
  \ifmaybe@ic % <= this is new
    \space
  \fi
}
\makeatother
\begin{document}
\title[Photoionisation Feedback in Molecular Clouds]{Photoionisation Feedback in a Self-Gravitating, Magnetised, Turbulent Cloud}
\author[S. Geen]
      {Sam Geen$^{1}$, Patrick Hennebelle$^{1}$, Pascal Tremblin$^{2,3}$, Joakim Rosdahl$^{4}$\\
{$^{1}$ Laboratoire AIM, Paris-Saclay, CEA/IRFU/SAp - CNRS - Universit\'e Paris Diderot, 91191, Gif-sur-Yvette Cedex, France}\\
{$^{2}$ Astrophysics Group, University of Exeter, EX4 4QL Exeter, UK}\\
{$^{3}$ Maison de la Simulation,  CEA-CNRS-INRIA-UPS-UVSQ,  USR 3441, Centre d’\'etude de Saclay, 91191 Gif-Sur-Yvette, France}\\
{$^{4}$ Leiden Observatory, Leiden University, P.O. Box 9513, 2300 RA, Leiden, The Netherlands}\\}
\date{\today}
\maketitle

% ABSTRACT  (☞ﾟヮﾟ)☞ 
\begin{abstract}
We present a new set of analytic models for the expansion of HII regions powered by UV photoionisation from massive stars and compare them to a new suite of radiative magnetohydrodynamic simulations of self-gravitating molecular clouds. To perform these simulations we use \simname{RAMSES-RT}, a Eulerian adaptive mesh magnetohydrodynamics code with radiative transfer of UV photons. \rev{In parallel to the simulations, we develop analytic models that describe the radial evolution of the HII region in a range of density and velocity fields. We give a radius at which the ionisation front should stop expanding (``stall''). If this radius is smaller than the distance to the edge of the cloud, the HII region will be trapped by the cloud. This effect is more severe in collapsing clouds than in virialised clouds, since the density increases dramatically over time, with much larger photon emission rates needed for the HII region to escape the cloud.} We also measure the response of Jeans unstable gas to the HII regions to predict the impact of UV radiation on star formation in the cloud. We find that the mass in unstable gas can be explained by a model in which the clouds are evaporated by UV photons, suggesting that the net feedback on star formation should be negative. \end{abstract}

% ------------------------------------------------------------------------------------------------------------------------------------------------------

\section{Introduction}
\label{introduction}

Massive stars form in dense gas clouds, where they heat the gas via ultra-violet (UV) photoionisation. The pressure difference created between these HII regions (where hydrogen is ionised from HI to HII) and the external medium causes an expansion wave that can disrupt the cloud in which the star formed and prevent (or enhance, via shock compression) further star formation. Previous theoretical papers by \cite{KahnF.D.1954,SpitzerLyman1978,Whitworth1979,Franco1990,Williams1997,Matzner2002} have successfully described the broad features of such regions using analytical arguments. These models have been successfully reproduced in hydrodynamic simulations of varying physical complexity by, e.g., \cite{Dale2005,Mellema2006,Krumholz2007a,Arthur2011,Walch2012,Walch2013,Dale2014}. Applying these models to observed HII regions as in \rev{\cite{Tremblin2014a,Didelon2015}} can help further our understanding of these systems, although there exists some degeneracy in the models that makes accurate comparisons difficult.

More recently, \cite{Hosokawa2006} and \cite{Raga2012} have modified the equations given by \cite{SpitzerLyman1978} for the thermal expansion of HII regions to include the shell inertia and external pressure terms respectively. Limits given by \cite{Garcia-Segura1996,Keto2002,Keto2003,Dale2012} put constraints on the expansion of HII regions due to pressure and accretion flows, leading to ultracompact HII regions that cannot escape the core or cloud in which they are formed. If the ram pressure and thermal pressure outside the region is equal to or larger than the thermal pressure in the photoionised gas, the front will be trapped.

The behaviour of ionisation fronts and HII regions is important because of their role in shaping the clouds in which they are formed. \cite{Mellema2006,Gritschneder2009,Tremblin2012,Walch2012} show in numerical simulations that HII regions can add a significant amount of momentum to the \ISM. Further, \rev{\cite{Rogers2013,Geen2015,Walch2015}} demonstrate that pre-supernova stellar feedback \rev{by stellar winds and photoionisation} allows supernova explosions to propagate much more efficiently than if they had exploded inside the dense cloud \citep[e.g.][]{Iffrig2015}. Star formation can also be quenched by UV photoionisation feedback, as demonstrated in \cite{Dale2011,Dale2012,Dale2013}. \cite{Walch2012a} note that a small increase in star formation rate can be achieved as the shock front around an expanding HII region compresses dense gas clumps in a molecular cloud, but the net effect of UV radiation from stars on star formation in a given cloud is expected to be negative.

Our goal in this paper is to use a combination of analytic and numerical work to introduce a new analytic model for the evolution of HII regions that \rev{describe} the interplay between self-gravity in clouds and UV photoionisation. We focus on the radius of the HII region with time, though this quantity is extended successfully to derivations of the momentum and mass of the HII region. Our model is based on previous analytic work but is extended to describe the photoionisation of turbulent molecular clouds. \rev{It considers the balance between the thermal pressure from UV photoionisation and ram pressure from turbulent or infalling gas in the external medium. A complete description of the dynamics of self-gravitating, magnetised clouds is beyond the scope of this paper. Instead, we adopt simple prescriptions based a power law density field with a velocity field equal to the escape velocity at each radius. We also provide models that sample directly the spherically-symmetric density and velocity profiles to compare to the simple power law model.} To test these models, we introduce a new suite of 3D simulations of UV photoionisation inside turbulent, magnetised self-gravitating clouds with sources of ionising radiation representing massive stars. In Section \ref{expansion} we present our models that will be applied to our simulations. We also (re-)derive equations for the limits to expansion that lead to ultracompact HII regions. In the Section \ref{methods} we introduce our suite of numerical simulations. Section \ref{overview} introduces the simulation results and addresses qualitatively the response of the cloud to UV photons and magnetic fields. In Section \ref{results:free} we compare our simulations to our analytic models assuming no external velocity field. In Section \ref{results:stalled} we include an external velocity field in our models and discuss at what point either model should be used. In Section \ref{results:feedback} we consider the expected response of star formation in the cloud to UV photoionisation. Finally we discuss the consequences of our results and scope for improvement in future work.

% ------------------------------------------------------------------------------------------------------------------------------------------------------

\section{Expansion of an HII Region}
\label{expansion}

\subsection{Basic Principles}
\label{expansion:basicprinciples}

The goal of this project is to better understand the evolution of HII regions in complex astrophysical environments. In order to interpret our results we require a theoretical understanding of the behaviour of HII regions in simpler environments that we can compare to our simulation results to determine at what point our results diverge from existing analytic work. The simulation code used in this paper, \simname{RAMSES-RT}, has been tested against other radiative transfer codes as well as analytic arguments by \cite{SpitzerLyman1978,Hosokawa2006} in an idealised environment and compares favourably to both \rev{\citep[see the Starbench project, presented in][]{Bisbas2015}}.

In this section we review our theoretical understanding of the evolution of ionisation fronts and derive expressions that will be used to model our results. An HII region is a volume of hydrogen gas that has been ionised, in this case by ultraviolet (UV) photons. Photoheated hydrogen has a temperature of approximately $10^4$ K. The precise temperature is regulated by the density and metallicity of the medium \citep[see, e.g.][]{Draine2011}. In this work we ignore the role of molecular hydrogen. Its dissociation energy of 4.5 eV means that there will be an extra energy cost in unbinding each H$_2$ molecule. In practice, however, our photons are normally much more energetic than the ionisation energy of hydrogen (13.6 eV), with the extra energy being lost to radiative cooling, and as such we do not expect accounting for this to change our results significantly.

The temperature difference between the photoionised and neutral hydrogen creates a pressure difference across the ionisation front, that triggers an expansion wave into the surrounding medium. The expansion of an HII region around a source of ionising UV photons is modelled in two phases \citep[see][]{KahnF.D.1954}. In the first phase, the front expands rapidly but hydrostatically to an equilibrium radius at which the emission rate of photons from the star is balanced by the recombination of photoionised hydrogen inside the ionisation front. This is called the \Stromgren radius, denoted by $r_s$. This phase is complete on the order of the recombination time in the medium, which at molecular cloud densities is on the order of kyrs. Since this is much shorter than the dynamical time of our system for our models we do not treat it in our models. Instead we assume that our front starts at the \Stromgren radius, which is typically small compared to the radius of the HII region at later times.

The total number of recombination events per unit time in a uniform, ionised sphere of radius $r_s$ dominated by hydrogen is given by $4\pi/3 r_s n_e n_0 \alpha_B$, where $n_e$ is the electron number density, $n_0$ is the initial hydrogen number density and $\alpha_B$ is the recombination rate in the photoionised gas. For a fully ionised gas, $n_e = n_0$. Hence we can equate this to the ionising photon emission rate $S_*$ and write  
\begin{equation}
 r_s = \left ( \frac{3}{4 \pi} \frac{S_*}{n_0^2 \alpha_B} \right )^\frac{1}{3}.
  \label{expansion:stromgren_radius}
  \end{equation} However, because this gas is at a higher temperature than the surrounding cloud material, the pressure inside the ionised gas is higher than the pressure outside and so the ionisation front expands. \cite{SpitzerLyman1978} makes the simplifying assumption that the ionisation front is coupled to the shock front \citep[see also][]{Dyson1980}. The ionisation front has a sound speed $c_i$, which is set by the radiative cooling equilibrium. In our simulations $c_i = 12.5~$km/s. This gives a constant temperature in the ionised gas of 8400 K \citep[based on][for an HII region at the same distance from the galactic centre as our Sun]{Tremblin2014a}. 

If we assume, as in \cite{SpitzerLyman1978}, that the ram pressure on the expanding front (left hand side) is equal to the thermal pressure inside the ionised gas (right hand side):

\begin{equation}
  n_{ext} \dot{r_i}^2 = n_i c_i^2,
  \label{expansion:ram_pressure}
\end{equation} where $n_{ext}$ is the hydrogen number density in the external medium just outside the shock radius, $r_i$ is the radius of the ionisation front and $n_i$ is the hydrogen number density in the ionised gas. 

As the ionisation front expands, the gas must remain in photoionisation equilibrium. Similarly to the arguments made for Equation \ref{expansion:stromgren_radius}, we can write

\begin{equation}
  r_{i} = \left ( \frac{3}{4 \pi} \frac{S_*}{n_i^2 \alpha_B} \right )^\frac{1}{3},
  \label{expansion:photon_balance}
\end{equation} where $r_{i}(t=0) = r_s$. In the next subsection we solve these two equations for a power law density profile.

\subsection{Expansion in a Power Law Profile}
\label{expansion:powerlaw}

\cite{SpitzerLyman1978} assumes a uniform density outside the HII region. \cite{Franco1990} instead use a power law density field with a flat core. We use a similar approach, but make the simplifying assumption that the core radius is negligible, since this does not significantly affect our results. Our density field is thus given by
\begin{equation}
n_{ext}(r) = n_0 \left ( \frac{r}{r_0} \right ) ^ {-w},
 \label{expansion:density_profile}
\end{equation} where $r$ is the distance from the centre of the cloud, and $r_0$ and $w$ are parameters fit to the cloud profile in our simulations, with $n_0$ the density at the position of the source, where $r_s$ is calculated. \rev{Fitting this power law to observed clouds is a non-trivial process, but it has been performed by, e.g. \cite{Yun1991,Bacmann2000,Nielbock2012,Didelon2015}. There are various difficulties in turning an observed 2D column density map into a full 3D density profile. A spherically symmetric approximation is easier to measure and calculate, however, than a full 3D density map. We discuss the validity of the assumption of spherical symmetry later in the paper.}

We substitute $n_{ext}$ in Equation \ref{expansion:ram_pressure} for its value in Equation \ref{expansion:density_profile}. We then substitute for $n_i$ in Equation \ref{expansion:photon_balance} and solve the differential equation in $\dot{r_i}$ to obtain

\begin{equation}
r_i(t) = \left( \frac{3}{4 \pi} \frac{S_*}{\alpha_B} \right )^\frac{\psi}{4} \left ( \frac{1}{n_0 r_0^w } \right ) ^\frac{\psi}{2} \left( \frac{1}{\psi} c_i t \right)^\psi
 \label{overview:radius}
\end{equation} where $\psi \equiv 4/(7-2w)$. The initial radius of the ionisation front is $r_s$, though this is typically much smaller than the radius of the front and thus can be neglected when calculating $r_i$ \citep[see also][]{Matzner2002}. For a homogeneous medium ($w$ = 0), we arrive at the \cite{SpitzerLyman1978} solution (assuming $r_s$ to be negligible). 

Similarly to \cite{Matzner2002}, we can calculate the properties of the HII region based on Equation \ref{overview:radius}. The mass in ionised gas is equal to the density in ionised gas $n_i$ multiplied by the volume of a sphere of radius $r_i$. We can use Equation \ref{expansion:photon_balance} to calculate this as
\begin{equation}
M_{ion} = \left( \frac{ 4 \pi}{3} \frac{ S_*}{\alpha_B}  \right ) ^ \frac{1}{2}  \frac{m_H}{X} r_i^{\frac{3}{2}} 
 \label{overview:ionmass}
\end{equation} where $m_H$ is the mass of one hydrogen atom and $X (= 0.76)$ is the hydrogen mass fraction in the gas. 

Note that this value is typically much smaller than the mass displaced by the ionisation front, since Equation \ref{expansion:photon_balance} evolves such that $n_i \propto r_i^{-3/2}$, so as the ionisation front expands $n_i << n_{ext}$, provided that the power law index $w$ is smaller than 3/2. This means that the mass of the shell is approximately the same as the mass of displaced gas, which can be obtained by integrating Equation \ref{expansion:density_profile}. The momentum of the shell is then the rate of change of $r_i$ multiplied by the mass of displaced gas:
\begin{equation}
M\dot{r} =  \frac{ 4 \pi}{3 - w}  r_0 ^ w n_0 \frac{m_H}{X}\frac{\psi}{t}r_i^{4 - w}
\label{overview:momentum}
\end{equation}

In subsequent sections we compare our simulation results to Equations \ref{overview:radius} to \ref{overview:momentum}, and determine at what point these expressions break down. In the following subsection we introduce theoretical arguments for why and where this should occur.

\subsection{Constraints on Expansion}
\label{expansion:raga}
 
The above expressions assume that the only force opposing the expansion of the HII region is ram pressure from the static neutral mass as the shock moves through it. In reality, there are a number of proceses that are able to resist the ionisation front's expansion. \cite{Garcia-Segura1996,Raga2012} argue that the pressure in the external medium can cause the ionisation front to stall, with \cite{Tremblin2014a} extending this model to the turbulence in the \ISM at the late stages of the front's expansion. As well as this, \cite{Keto2002} gives some theoretical limits at which accretion must block the expansion of HII regions in stellar cores, while \cite{Dale2012} states that for molecular clouds if the escape velocity at the \Stromgren radius exceeds the sound speed in the photoionised gas, the front remainds trapped.

We now adapt the model given in the previous section to determine the point at which it breaks down in the presence of an external gas infall, thermal pressure or turbulence. Equation \ref{expansion:ram_pressure} can be modified to include terms in turbulence and infall velocity \citep[derived in Appendix \ref{infalllimited}, based on][]{Raga2012}:

\begin{equation}
\frac{1}{c_i}\frac{\mathrm{d}r_i}{\mathrm{d}t} = F(r,t)-\frac{c_{ext}^2}{c_i^2}\frac{1}{F(r,t)}+\frac{v_{ext}(r,t)}{c_i} ,
\label{expansion:raga_like}
\end{equation} where
\begin{equation}
F(r,t) \equiv \sqrt{\frac{n_i}{n_{ext}}} = \left(\frac{r_{s}}{r_i}\right)^{3/4}\left(\frac{n_0}{n_{ext}(r,t)}\right)^{1/2} .
 \label{expansion:raga_term}
 \end{equation} $c_{ext}$ is a term including the sound speed and turbulent motions in the gas just outside the shock radius and $v_{ext}$ is the velocity of the gas just outside the shock radius normal to the shock surface (assumed in 1D to be radial from the source position).

In the limit where $c_{ext}\rightarrow 0$ and $v_{ext} \rightarrow 0$, we arrive at the solution in Equation \ref{overview:radius}. For a general case where $c_{ext}$ and $v_{ext}$ are non-negligible, it is not possible to solve Equation \ref{expansion:raga_like} analytically. However, we can make arguments concerning the limit at which $\dot{r_i} = 0$. This is the radius at which the ionisation front will stall and be unable to expand further. In this limit for infall-dominated flows, $F(r,t) \simeq \frac{v_{ext}(r,t)}{c_i}$. Similarly, for a cloud dominated by thermal pressure, $F(r,t) \simeq \frac{c_{ext}}{c_i}$. \cite{Tremblin2014a} use the approximation $c_{ext} = 1/3 \sigma^2$ where $\sigma$ is the velocity dispersion of the turbulence in the external medium. If the velocity of the gas grows over time, the ionisation front can even shrink as the HII region is crushed by accretion.

For a quasi-static power law density profile as given in Equation \ref{expansion:density_profile} where $r_i = r_s$ at $t=0$, we can write 
\begin{equation}
F(r,t) = \left(\frac{r_{s}}{r_i}\right)^{\frac{3}{4}} \left(\frac{r_{0}}{r_i}\right)^{-\frac{w}{2}} .
\label{expansion:Fpowerlaw}
\end{equation} The radius at which the ionisation front stalls ($\dot{r_i} \rightarrow 0$) is found by substituting Equation \ref{expansion:Fpowerlaw} into Equation \ref{expansion:raga_like} for a flow dominated by infall, with an identical expression in $c_{ext}$ if the sound speed in the neutral gas dominates:
\begin{equation}
r_{stall} = r_0 \left( \frac{r_s}{r_0} \right)^{\frac{3}{3 - 2w}}\left( \frac{c_i}{v_{ext}} \right)^{\frac{4}{3 - 2w}}
\label{expansion:stall_radius}
\end{equation} where $r_{stall}$ is the value of $r_i$ at which $\dot{r_i}=0$. If $w \simeq 3/2$, then the front cannot expand beyond $r_0$ regardless of radius if the escape velocity exceeds the sound speed.

Alternatively, we can write

\begin{equation}
r_{stall} = \left(\frac{3}{4 \pi}\frac{S_*}{\alpha_B} \right ) \left(\frac{m_H}{X} \frac{c_i^2}{P_{ram}}\right)^{2/3},
 \label{expansion:pressure_balance}
\end{equation} where $P_{ram}$ is the ram pressure on the shock from an external velocity field ($\frac{m_H}{X} n_{ext} v_{ext}^2$), turbulence or thermal pressure ($\frac{m_H}{X} n_{ext} c_{ext}^2$). Note that $P_{ram}$ and $v_{ext}$ can depend on radius (as we discuss in the following section), in which case the dependence of $r_{stall}$ on $c_i$ and $S_*$ will change accordingly.

\subsection{Stalled Expansion in an Accreting or Virialised Cloud}
\label{expansion:virialised}

The infall velocity used to calculate these limits depends on the dynamics of the cloud. A complete treatment of this subject is beyond the scope of this paper. \cite{Hunter1962} gives a simple spherical collapse model in which a cloud accretes onto the centre at free-fall speeds. In our work, the cloud is supported by turbulence, as well as thermal and magnetic pressure. For a cloud in virial equilibrium, where $\frac{1}{2}v_{ext}^2 = \frac{3}{5}\frac{GM}{R}$ (G is the gravitational constant, M is the mass of the cloud and R is the radius of the cloud), we can estimate the velocity of the gas inside the cloud to be approximately $\sqrt{\frac{3}{5}}$ the escape velocity, though this is a crude approximation to the actual velocity structure, which is continuously evolving. In addition these motions will be random, rather than directed solely towards the centre of the cloud. One source of infall in the cloud is from relaxation processes. As the turbulence inside the cloud decays, the cloud loses turbulent support and collapses inwards \citep[see, e.g., ][]{Gao2015}. This is complicated by fragmentation, in which dense clumps form and cause the structure of the cloud to become highly non-spherical. Since a full model of the dynamics of a cloud is beyond the scope of the current paper, for the rest of the paper we simply assume $v_{ext} = v_{esc}$. 

Using the power law density field given in Equation \ref{expansion:density_profile} and the fact that $v_{esc}^2 = 2 G M(<r) / r$ for $M(<r)$ being the mass enclosed inside $r$, we can compute the escape velocity of the cloud as
\begin{equation}
v_{esc}^2 = 8 \pi G n_0 \frac{m_H}{X} r_0^w \frac{r^{2-w}}{3-w}.
\label{virialised:vesc}
\end{equation} An equivalent expression for the ram pressure is
\begin{equation}
P_{ram} = \frac{8 \pi G}{3-w} \left(\frac{m_H}{X} n_0 r_0^w r^{1-w}\right)^2.
\label{virialised:pram}
\end{equation} Applying this to Equation \ref{expansion:pressure_balance}, where $r = r_{stall}$, we can write
\begin{equation}
r_{stall} = \left( \frac{3}{4 \pi}\frac{S_*}{\alpha_B} \right )^\frac{\theta}{4} \left( \frac{X}{m_H} \frac{(3-w)}{8 \pi G} \right )^\frac{\theta}{2} \left( \frac{c_i}{n_0 r_0^{w}} \right )^\theta
 \label{virialised:rstall}
\end{equation} for a power law density field with gas moving radially inwards at the escape velocity, where $\theta \equiv 4/(7-4w)$.

In Section \ref{results:stalled} we compare this model to our simulation results to determine whether it holds in a 3D density field.

% METHODS
\section{Methods}
\label{methods}

\subsection{Numerical simulations}
\label{methods:numsim}

%TABLE OF SIMULATIONS AND THEIR PARAMETERS

\begin{table}
\begin{tabular}{l c c c c c c c c c}
   \textbf{Name} & \textbf{log$_{10}$($S_{*}$/$s^{-1}$)} & \textbf{B-field?} & \textbf{$t_{ff}$/Myr} \\
  \hline
  \textbf{Varying UV Source} \\
  \hline
 \simname{N00\_B02} & (no photons) & \tick & 1.25 \\
 \simname{N47\_B02} & 47 & \tick & 1.25 \\
 \simname{N48\_B00} & 48 &       & 1.25  \\
 \simname{N48\_B02} & 48 & \tick & 1.25  \\
 \simname{N49\_B02} & 49 & \tick & 1.25  \\
  \hline
  \textbf{Delayed UV Emission} \\
  \hline
 \simname{N48\_B02\_F2} & 48 & \tick & 1.25  \\
 \simname{N48\_B02\_F3} & 48 & \tick & 1.25  \\
  \hline
  \textbf{Varying Compactness} \\
  \hline
 \simname{N00\_B02\_C} & (no photons) & \tick & 0.156 \\
 \simname{N48\_B02\_C} & 48 & \tick & 0.156 \\
 \simname{N00\_B02\_C2} & (no photons) & \tick & 0.527 \\
 \simname{N48\_B02\_C2} & 48 & \tick & 0.527  \\
  \textbf{Infall-Dominated} \\
  \hline
 \simname{N00\_B00\_IF} & (no photons) &  & 1.25 \\
 \simname{N48\_B00\_IF} & 48 &  & 1.25 \\
 \simname{N49\_B00\_IF} & 49 &  & 1.25 \\
 \simname{N50\_B00\_IF} & 50 &  & 1.25 \\
  \hline
\end{tabular}
  \caption{Table of the simulations included in this paper. ``\simname{N}'' refers to the photon emission rate $S_*$ in all frequency groups in photons per second as a power of 10 (with ``\simname{N00}'' referring to a zero photon emission rate).  ``\simname{B02}'' refers to a ratio between the free-fall time and Alfven crossing time of 0.2, while ``\simname{B00}'' refers to a uniform field of 0$~\mu$G. ``\simname{F2}'' and ``\simname{F3}'' refer to simulations where the source is turned on at 2 $t_{ff}$ and 3 $t_{ff}$ respectively (rather than after 1 $t_{ff}$ as in all other simulations). \rev{``\simname{IF}'' refers to an ``infall-dominated'' setup in which there is no turbulence in the initial conditions, so the cloud undergoes radial collapse. These runs are spherically symmetric due to the absence of turbulence, which allows better comparison to our 1D analytic models.} All simulations are run with the ``Fiducial'' cloud initial conditions except those labelled ``\simname{C2}'' (``More Compact'') and ``\simname{C}'' (``Most Compact''). The Fiducial, More Compact and Most Compact clouds have ratios of sound crossing time $t_{sct}$ to $t_{ff}$ of 0.1, 0.075 and 0.05 respectively. See Section \ref{methods:numsim} for more details.}
\label{methods:numsimtable} 
\end{table}

We now introduce the numerical simulations designed to test the analytic expressions given in the previous section. The properties of each of the simulations run are given in Table \ref{methods:numsimtable}. In all the simulations in this paper we use a cloud of $10^4$ solar masses. We leave an analysis of different cloud masses to future work, though our analytic results should apply to clouds of different masses. All of our simulations use a cubic adaptive octree mesh with a coarse spatial resolution $256^3$ cells, or 8 levels of the octree, and two additional levels of refinement, leading to a total of 10 levels of refinement and a maximum effective spatial resolution of $1024^3$ cells in the most refined cells. We refine based on the Jeans criterion, such that if the thermal energy of a cell above the maximum level of refinement is lower than its gravitational potential, we trigger an additional level of refinement. We implement outflow boundary conditions, such that any matter that leaves the simulation volume is assumed to be lost to the system.

\subsection{Initial Conditions}
\label{methods:cloud}

\begin{figure}
 \centerline{\includegraphics[width=0.98\hsize]{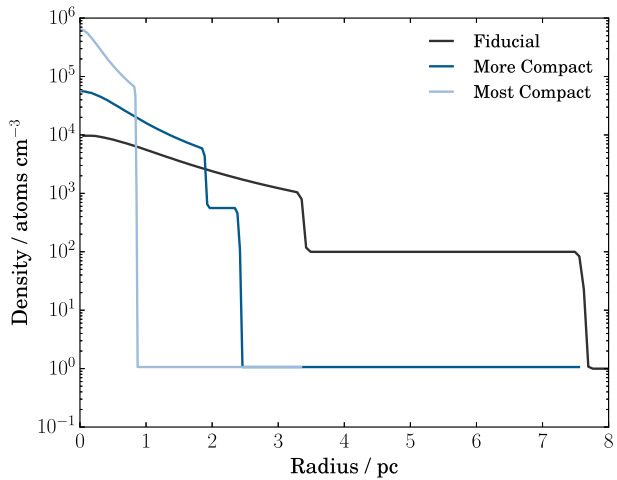}}
 \caption{Initial hydrogen number density against radius for each of the cloud profiles (Fiducial, More Compact and Most Compact) used in the simulations. The density field in the Fiducial case extends to 13.5 pc (the edge of the simulation volume). See Section \ref{methods:cloud} for details.}
  \label{methods:initialdensities}
 \end{figure}

The setup for our simulations is similar to that in \cite{Iffrig2015}. In Figure \ref{methods:initialdensities} we plot initial hydrogen number density profiles for each of the set of initial conditions used in this paper. We impose a spherically symmetric cloud onto the simulation volume with an inner hydrogen number density profile given by a pseudo-isothermal sphere $n_{ext}(r,t=0) = n_0 / (1 + (r/r_c)^2)$. Most of the simulations in this paper are run using the ``Fiducial'' cloud, which has $n_0 = 9370$ \atcc and $r_c = 1.12$ pc, with a free-fall time of 1.25 Myr, defined as $\sqrt{\frac{3 \pi}{32 G \rho}}$, where $\rho$ is the mean cloud density in the pseudo-isothermal portion of the cloud. We impose a cut-off at $3~r_c$ (where $n_{ext} = 0.1~n_0$). The temperature inside the inner part of the cloud is set to 10 K. Outside this region we impose a flat density field at 93.7 \atcc out to 7.6 pc. We then impose a medium outside this radius with hydrogen number density 1 \atcc at a temperature of 4000 K. The total box size is 27 pc, giving a spatial resolution of 0.1 pc on the coarse grid and a maximum resolution of 0.026 pc in the most refined cells. The centre of the cloud is located at the centre of the simulation volume. For simulations containing a magnetic field, the field is imposed so that the ratio between the free-fall time and the Alfven crossing time is 0.2 (denoted in the simulation names by \simname{B02}), corresponding to a value of B = 20 $\mu$G at the centre of the sphere and 4.4 $\mu$G at the edge. In addition to this, we run one control simulation with no magnetic field. A turbulent velocity field is imposed over the grid, such that the kinetic energy in turbulence in the cloud is approximately equal to the gravitational energy of the cloud. The turbulence has a Kolmogorov power spectrum with random phases.

We also run two similar simulations in denser environments in order to probe the effect of cloud density on the ionisation front. The ``More Compact'' (suffix \simname{\_C2}) and ``Most Compact'' (suffix \simname{\_C}) clouds have ratios of sound crossing time to free-fall time of 0.75 and 0.5 of the fiducial setup respectivey. This gives free-fall times of $0.75^3$ and $0.5^3$ of the free-fall time in the fiducial cloud (0.527 and 0.156 Myr) and cloud radii of $0.75^2$ and $0.5^2$ respectively, based on the expression for the free-fall time given above. The Most Compact cloud has an initial density profile with the same equation for $n_{ext}$ as above but with $n_0 = 6\times10^5$ \atcc and an edge radius of $3~r_c= 0.8$ pc. The More Compact cloud has $n_0 = 5.3\times10^4$ \atcc with an edge radius of $3~r_c = 2.0$ pc, with a flat density field of 530 \atcc out to 2.5 pc. As in the fiducial setup we include a diffuse medium surrounding the cloud with density 1 \atcc and temperature 4000 K. The More Compact runs have a maximum spatial resolution of 0.015 pc and the Most Compact runs 0.0066 pc.

\rev{We run a further four runs using the Fiducial cloud, except with no initial velocity field and a magnetic field strength set to zero, causing the cloud to be infall-dominated and spherically symmetric. These simulations have the suffix ``\simname{IF}''. We do this to test our model in a spherically symmetric environment. We include sources of log$_{10}(S_*)$ = 48,49 and 50 as well as a run with no source of photons.}

\subsection{UV Source Properties}
\label{methods:sources}

For each cloud density we run a simulation without photons (\simname{N00\_B02}, \simname{N00\_B02\_C2} and \simname{N00\_B02\_C}). We also run a set of simulations with a constant source of photons placed at the centre of the simulation volume. For each cloud density we run a simulation in which an emission rate of ionising photons $S_* = 10^{48}$ s$^{-1}$ is turned on after one free-fall time (\simname{N48\_B02}, \simname{N48\_B02\_C2} and \simname{N48\_B02\_C}, with $t_{ff} =$ 1.25, 0.527 and 0.156 Myr respectively). This source is modelled as a blackbody similar to a B0V star with surface temperature 33700 K, radius 8.3 \Rsolar and mass 20 \Msolar \citep[based on][ calibrated to give roughly $10^{48}$ ionising photons per second]{Sternberg2003}. For each photon group we sample this blackbody and add the appropriate photon emission rate to each group. In every simulation in this paper we assume that the sources have a constant emission rate, though massive stars will in practice have a decreasing emission rate in time as their envelopes expand and decrease in temperature \citep[as we model in ][]{Geen2015}. However, \cite{Sternberg2003} give a more or less constant photon emission rate for 3-5 Myr, depending on the lifetime of the most massive star in the cluster, which is as long or longer than each of the simulations in this paper.

For the Fiducial cloud we run two further simulations identical to \simname{N48\_B02} but with different photon emission rates - a $10^{47}$ photons/s source modelled on a B1V star with surface temperature 28500 K, radius 4.7 \Rsolar and mass 12 \Msolar \citep[based on][]{Pecaut2012}, and a $10^{49}$ photons/s source modelled on 05V star with surface temperature 46000 K, radius 11.2 \Rsolar and mass 55 \Msolar \citep[based on][]{Vacca1996}. These simulations are labelled \simname{N47\_B02} and \simname{N49\_B02} respectively. As in simulation \simname{N48\_B02} we turn on the source after one free-fall time in the Fiducial cloud, 1.25 Myr. The goal of this is to test the expansion of the ionisation front in the cloud for various photon emission rates. In principle higher photon emission rates exist for very massive stars or more massive clusters, though we find that our $10^{49}$ photons/s star can easily destroy the cloud and thus any sources above this value are expected to behave similarly in runs with a turbulent cloud. For the infall-dominated runs, where the density of the cloud increases dramatically without turbulent support, we include an extra $10^{50}$ photons/s source. This a blackbody based on a 120 \Msolar star from \cite{Schaller1992} with surface temperature 56000 K and radius 22.2 \Rsolar.

We also run a simulation identical to \simname{N48\_B02} but without a magnetic field in order to study the effect that the field has on the cloud and HII region. We label this simulation \simname{N48\_B00}.

In addition, we run a further two simulations that are identical to \simname{N48\_B02} except that we delay the time the source is turned on. These are labelled \simname{N48\_B02\_F2} (where the source is turned on at 2 $t_{ff}$ = 2.5 Myr) and \simname{N48\_B02\_F3} (where the source is turned on at $3~t_{ff}$ = 3.75 Myr). We do this in order to test whether allowing the cloud to evolve further has an impact on the resulting HII region. This is motivated by the fact that the cloud structure changes dramatically over $3~t_{ff}$, and we wish to determine whether this has an effect on the subsequent expansion of the HII region.

\subsection{Simulation Setup}
\label{methods:simulations}

The initial conditions are evolved using a modified version of the radiative magnetohydrodynamics code \textsc{RAMSES-RT} \citep{Teyssier:2002p533,Fromang2006,Rosdahl2013}. \textsc{RAMSES-RT} uses a first-order moment method for the advection of photons, closing the set of equations with the local M1 expression for the radiation pressure tensor. It tracks the ionisation states of hydrogen and helium in the gas, and couples the interactions between the photons and the gas on-the-fly. We split the radiation into three groups, bracketed by the ionisation energies of HI, HeI and HeII (13.6, 24.6 and 54.2 eV for the lower bounds of each), though in practice helium ionisation is found to have less of an impact than hydrogen ionisation on our results. More energetic photons do not tend to significantly alter the results since the temperature of the ionised gas is determined largely by the cooling of photoionised metals in the HII region. In all of the simulations in this paper we assume a Solar metallicity everywhere at all times. We do not consider photons below the ionisation energy of hydrogen, nor do we include radiation pressure, which we reserve for future work. A reduced speed of light of $10^{-4}$ c (= 30 km/s, or 2.4 $c_i$) is used. We do this because the speed of light sets the Courant factor in the timestep calculation \citep[see][for details]{Rosdahl2013}, and thus a reduced speed of light improves the efficiency of our simulations dramatically. We chose the minimum value such that the speed of ionisation fronts in our simulations is calibrated to be the same as that for a larger speed of light.

Gas above a hydrogen ionisation fraction of 0.1 is considered to be photoionised, and is set to a temperature of 8400 K, following \cite{Tremblin2014a} for an HII region at the same distance from the galactic centre as the Sun. This is based on observed HII regions, taking the galactic radius of our cloud to be at the same distance from the galactic centre as the Sun. For gas below this ionisation fraction, non-equilibrium cooling on hydrogen and helium is calculated based on the prescription given in \cite{Rosdahl2013}. Radiative cooling of metals and background heating from the \ISM are calculated according to an analytic cooling function given in \cite{Iffrig2015}. This function is a combination of the low-temperature cooling function of \cite{Audit2005} and the high temperature component of \cite{Sutherland1993}, giving a cooling and heating function similar to that used by \cite{Joung2006}.
 
% ------------------------------------------------------------------------------------------------------------------------------------------------------

\section{Expansion in a Spherically Symmetric Collapsing Cloud}
\label{infall}

\rev{In the first instance we compare our models to simulations in a spherically-symmetric density field without turbulence. This is to allow more direct comparison with our 1D models. We do not run the simulations in 1D as \textsc{RAMSES} does not support spherical coordinates, and thus a direct 1D comparison between our code and the model is not possible. \textsc{RAMSES-RT} has taken part in the \simname{STARBENCH} code comparison project \citep{Bisbas2015} and has been shown to agree with other codes and analytic theory for the expansion of HII regions in a selection of static media, including tests that compare our results to the equations of \cite{Hosokawa2006} and \cite{Raga2012}. See \cite{Rosdahl2013} for additional analytic model comparisons.}

\rev{In order to bridge the gap between these static media and a fully turbulent cloud, we compare our equations to a source expanding in a cloud without turbulence. It should be noted that a cloud with no turbulent support is considered to be unlikely. Observations by, e.g. \cite{Peretto2013}, find that clouds tend to be globally virialised. This turbulence-free simulation is thus provided largely for the sake of model comparison.}

\rev{In Simulations \simname{N00\_B00\_IF}, \simname{N48\_B00\_IF}, \simname{N49\_B00\_IF} and \simname{N50\_B00\_IF}, we run the Fiducial cloud without an initial turbulence field or magnetic field but with self gravity (see Section \ref{methods}). In this case the density of the cloud increases dramatically on a timescale of the freefall time \citep[see][]{Larson1969}. We do not provide a detailed algebraic solution for this case. Instead, we compute a ``non-static'' solution to Equation \ref{expansion:raga_like} that includes the infall velocity of the gas at each radius. We do this by sampling the time-dependent radial density and velocity field in Simulation\simname{N00\_B00\_IF} and compare that to our simulation results. In other words, we use the simulation without feedback to provide the density and infall velocity for the model. This allows us to test the accuracy of the front propagation equation in the absence of a detailed theoretical model for the radial collapse of the cloud.}

\begin{figure}
 \centerline{\includegraphics[width=0.98\hsize]{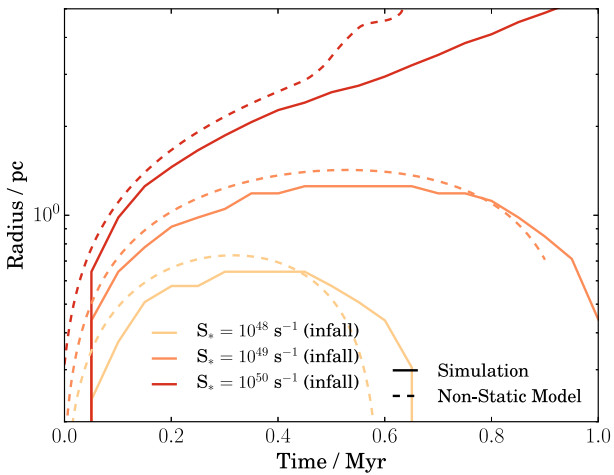}}
 \caption{\rev{The median radius of the ionisation front over time compared to Equation \protect\ref{expansion:raga_like} in Simulations \simname{N48\_B00\_IF}, \simname{N49\_B00\_IF} and \simname{N50\_B00\_IF}. The solid lines show the median radius of the ionisation front in each simulation. The dashed lines show the ``Non-Static Model'' solution to Equation  \protect\ref{expansion:raga_like}, which includes an external velocity field.}}
  \label{infall:ragacompare}
 \end{figure}

\rev{In Figure \ref{infall:ragacompare} we plot this comparison. Our solution to Equation \ref{expansion:raga_like} (dashed line) follows the evolution of the simulation results (solid line). There is a short time lag between our solutions and the simulation results. This is because the velocity field of the cloud responds to the HII region, so there is some divergence between the simulation without photons used to calculate the analytic solution and the simulated clouds containing an HII region. In addition, the simulation outputs have limited time resolution, and the velocity and density fields are linearly interpolated in time. Nonetheless, we reproduce the form of the simulation results, in particular the parabolic arc of the radius over time as the increasing density crushes the ionisation front. We also find agreement that beyond a certain photon flux the front is able to escape the cloud, as we discuss in more detail in Section \ref{results:stalled} for the turbulent clouds. In particular, we find that the photon emission rate needed to escape the cloud is much higher in the infalling cloud than in the turbulent cloud because of the dramatic increase in density over time due to the collapse of the cloud, which lacks support against gravity from turbulence.}

\section{Simulation Overview}
\label{overview}

We now review the qualitative properties of simulations in the Fiducial cloud and the effect of the magnetic field on the structure of the cloud and the HII region. \rev{We leave quantitative comparison of the simulation results to our analytic models to subsequent sections}.

\subsection{Before the First Star}
\label{overview:before}

\rev{In this section we briefly review the behaviour of the cloud with and without the magnetic field in the absence of photons as it pertains to the subsequent evolution of the HII region. After $t=0$, the cloud begins fragmenting under the influence of its initial turbulence, magnetic fields and self-gravity. This fragmentation occurs over $\sim t_{ff}$. In the absence of a source of photons, these clumps move towards the centre over a further $2~t_{ff}$. Since the cloud is virialised, this infall is due to mass segregation and turbulent decay. The cloud is distorted from its initial spherical shape, though maintaining the sharp discontinuity between the cloud gas at above 100 \atcc and the external medium at 1 \atcc. The shape of the cloud is important insofar as it determines how far the HII region has to expand to escape the cloud. At $t_{ff}$, we sample the distance from the centre of the cloud to the edge along evenly sampled lines of sight (see Appendix \ref{appendix:lossampling}). We find a roughly uniform probability distribution of radii from 3 to 12 pc in the Fiducial cloud. Thus the HII region must travel at least 3 pc to escape the cloud, and if it travels 12 pc it will have completely disrupted the cloud.}

\begin{figure*}
 \centerline{\includegraphics[width=0.48\hsize]{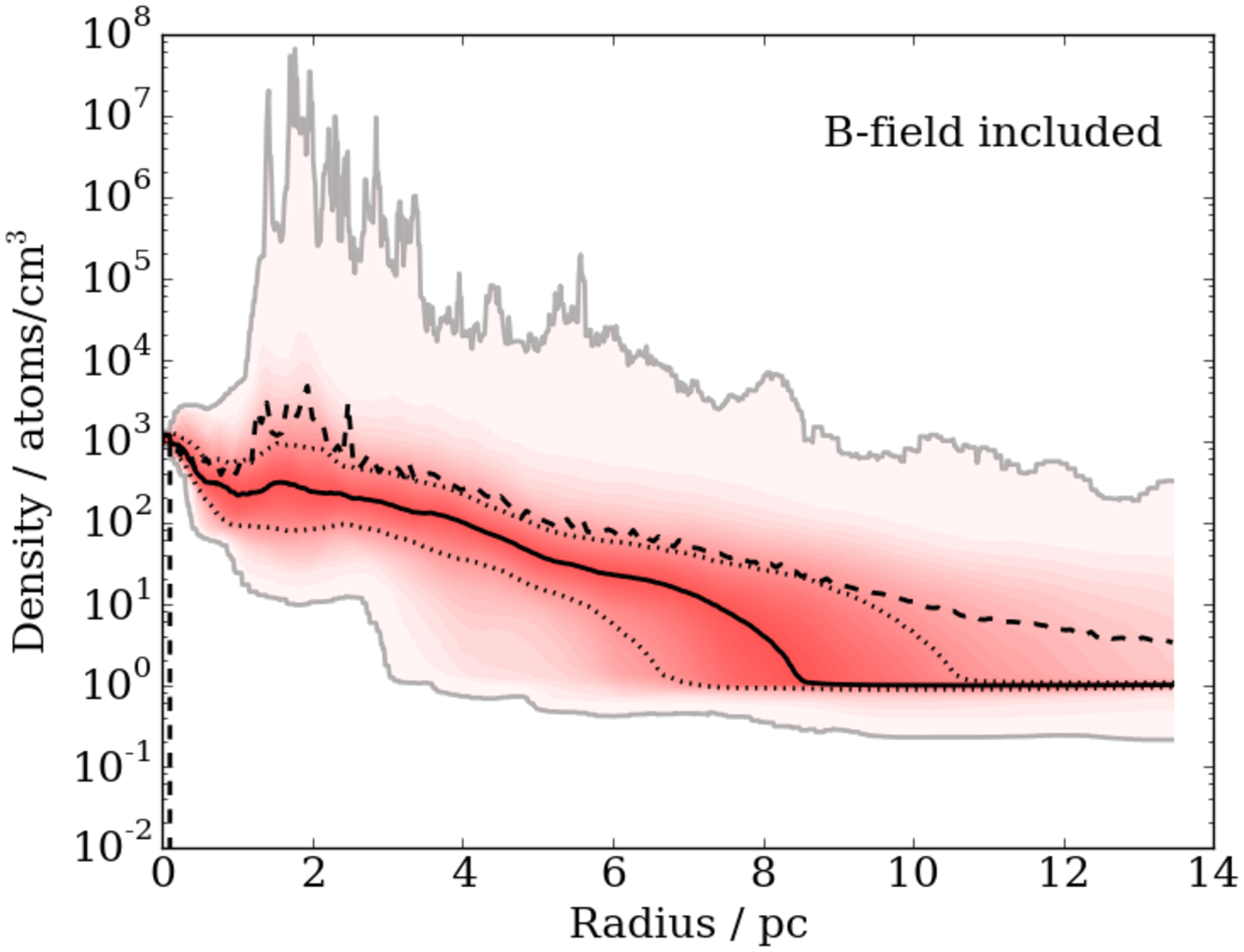} \includegraphics[width=0.48\hsize]{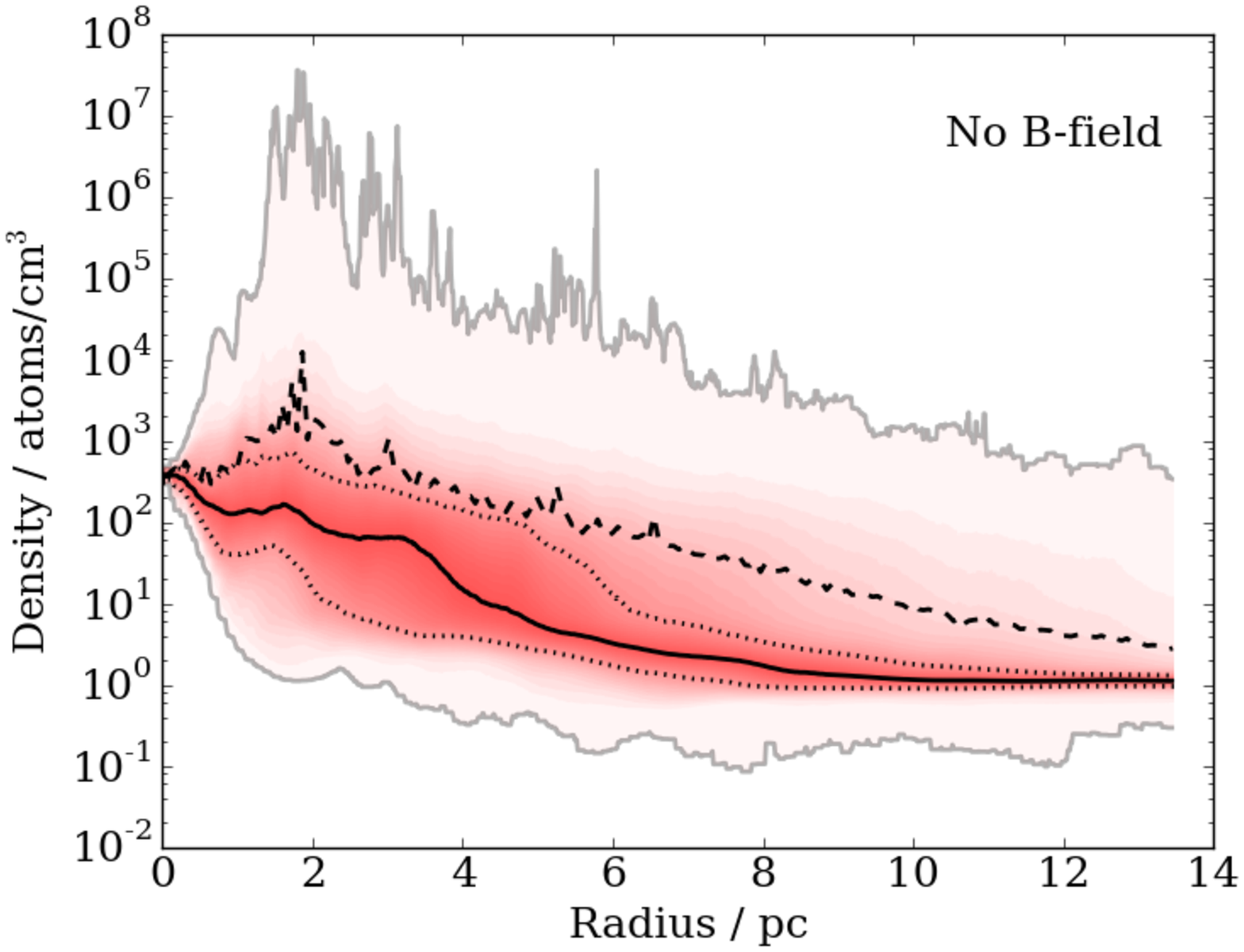}}
 \centerline{\includegraphics[width=0.48\hsize]{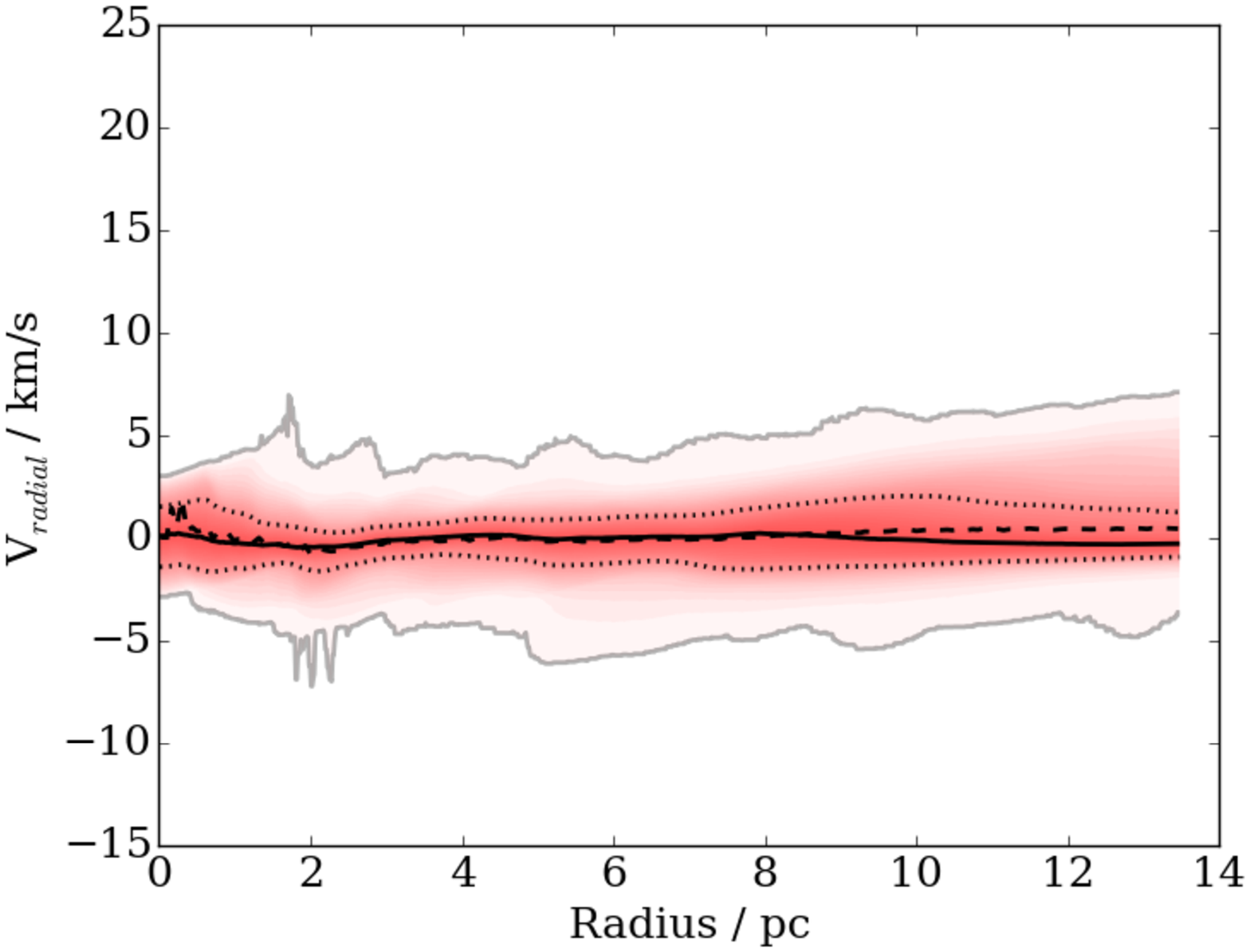} \includegraphics[width=0.48\hsize]{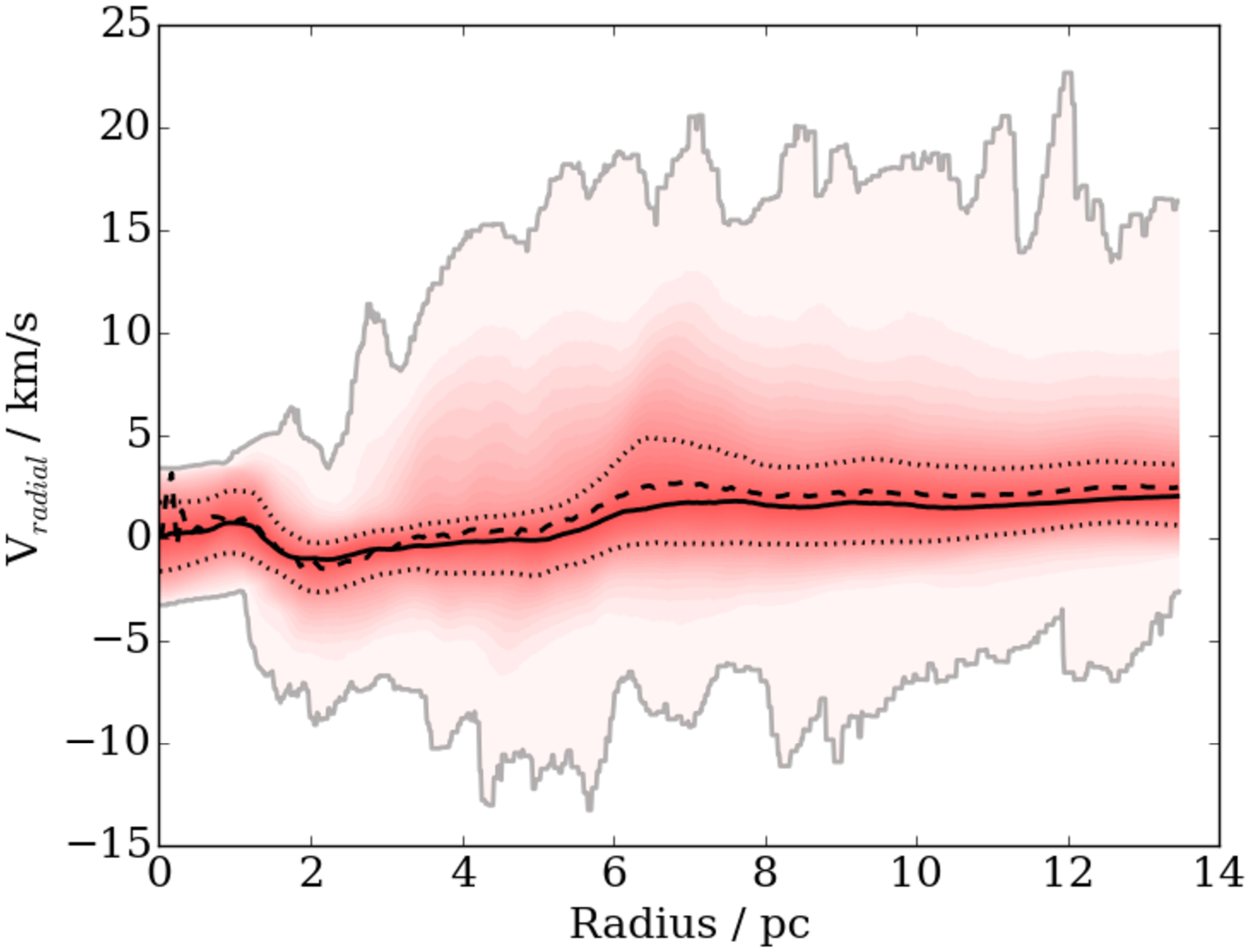}}
 \caption{Probability distribution in density and radial velocity at each radius in simulations with and without a magnetic field. The top row shows hydrogen number density, while the bottom row shows the radial velocity, where positive values are flows away from the centre. The left-hand figures show the fiducial cloud at $t_{ff}$ with a magnetic field (simulation \protect\simname{N48\_B02}) while the figures on the right are taken at the same time from an identical setup but without a magnetic field (\protect\simname{N48\_B00}), noting that neither simulation contains a source of photons before $t_{ff}$ = 1.25 Myr. In each plot we sample the density and radial velocity along evenly spaced lines of sight from the source position (see Appendix \protect\ref{appendix:lossampling}). The grey lines denote the maximum and minimum values for all lines of sight at each radius. The solid black line shows the (volume-weighted) median value at each radius. The dashed black line is the mean value at each radius. The dotted black lines show the values at the 25th and 75th percentile. The shaded region goes from white (maximum or minimum at each radius) to red (median).}
  \label{overview:profiles}
 \end{figure*}

\rev{In Figure \ref{overview:profiles} we plot the cloud structure $t_{ff} = 1.25~$ Myr in simulations \simname{N48\_B02} and \simname{N48\_B00}, i.e. the Fiducial cloud with and without magnetic fields. At this time no UV photons have been emitted. In this figure we sample profiles along lines of sight as before (Appendix \ref{appendix:lossampling}) and plot the (volume-weighted probability) distribution of density and radial velocity at each radius as a colour gradient. The grey lines denote the maximum and minimum values at each radius, and the black line the median value at each radius. The red values represent probability bins around the median at each 2nd percentile, and the white values the highest and lowest percentiles.}

\rev{The velocity field is relatively flat in the magnetised cloud, whereas in the non-magnetised case flows up to four times faster are seen in the non-magnetised cloud. This is because, in the absence of magnetic support the cloud must rely on support from turbulence. The mean density profiles (dashed lines in Figure \ref{overview:profiles}) with and without magnetic fields are very similar. Fitting the mean density profiles of both clouds at $t_{ff}$ to a power law, we find power law indexes ($-w$ in Equation \ref{expansion:density_profile}) of -0.74 in the cloud containing a magnetic field and -0.6 in the cloud with no magnetic field.}

\begin{figure*}
 \centerline{\includegraphics[width=0.98\hsize]{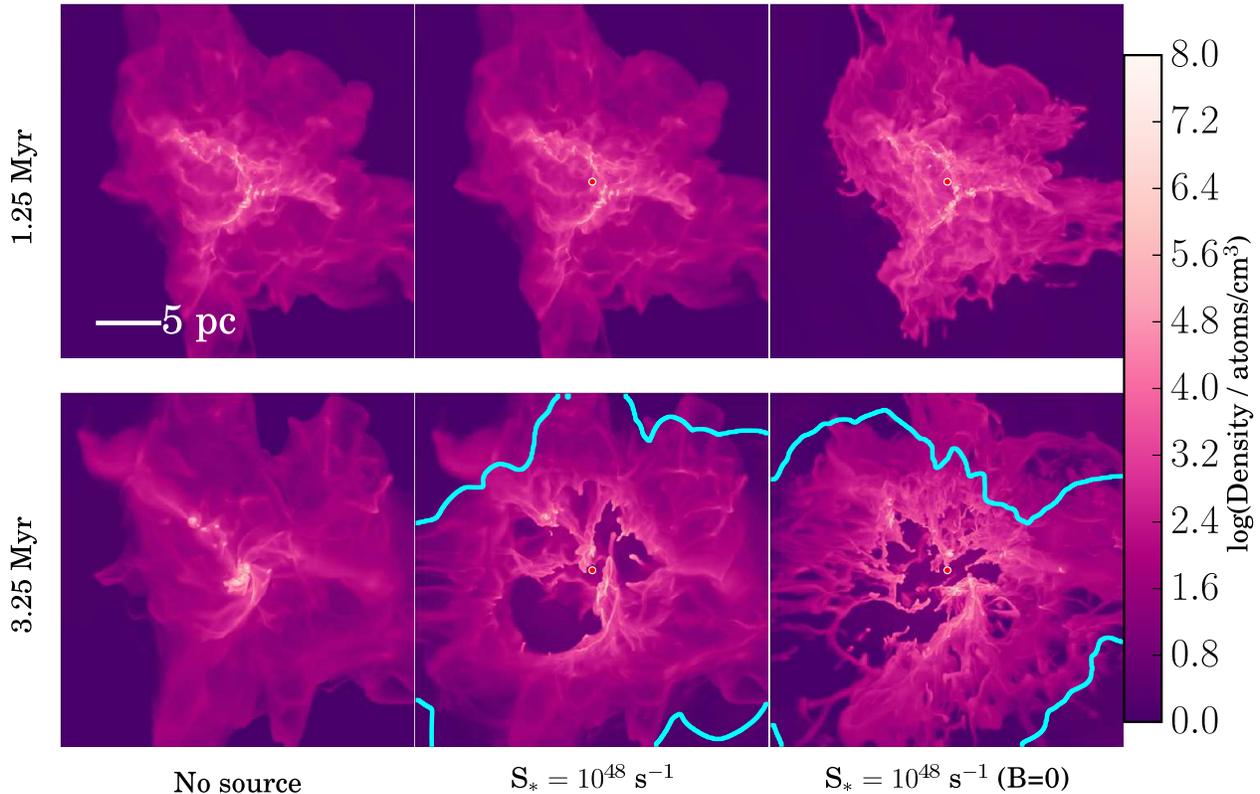}}
 \caption{Sequence of projections of the ``fiducial'' cloud showing maximum hydrogen number density along the line of sight. The cyan (light blue) contour marks the edge of the HII region (measured as a hydrogen ionisation fraction above 0.1) when projected onto the image. A red circle indicates the position of the UV source if one is included in the simulation. The left column is simulation \simname{N00\_B02}, the middle column is \simname{N48\_B02}, and the right-hand column is \simname{N48\_B00}. The top row shows each simulation at 1.25 Myr (1 $t_{ff}$) and the bottom column at 3.25 ($t_{ff}$ + 2 Myr). All images show the full cubic simulation volume of length 27 pc. The presence of a source of UV photons dramatically alters the overall structure of the cloud, while the magnetic field alters the filamentary structure and the shape of the HII region.}
  \label{overview:images}
 \end{figure*}

\rev{However, there are differences in the angular distribution of matter as shown in the median (solid black line) and interquartile ranges (dotted black lines). These drop off more quickly in the simulation without magnetic fields. Since the mean density profile is the same, it leads us to conclude that more mass is found in small clumps in the non-magnetised cloud. This agrees with the findings of \cite{Hennebelle2013} - see \cite{Soler2013} for a discussion. This clumping can also be seen in the top panel of Figure \ref{overview:images}, though it is clearer after the HII region has formed, which we discuss in the next section.}

\subsection{HII Regions with and without Magnetic Fields}
\label{overview:expansion}

\rev{Once the source is turned on at $t_{ff}$, the ionisation front begins to expand. The densest clumps remain embedded while the less dense gas is pushed away. There is a competition between the acceleration of the clumps by the rocket effect \citep[see][]{Oort1955}, in which UV photoevaporation from the surface closest to the source causes the clump to accelerate away, and the effects of gravity. The most massive clumps, unless resisted by the UV photons, will tend to move further inwards due to mass segregation \citep{Spitzer1969}.}

\rev{The increased fragmentation of the cloud when no magnetic field is present provides more channels of low-density gas between the dense clumps. This in turn allows the HII region to escape preferentially through these channels. This can be seen in the bottom panels of Figure \ref{overview:images}, where we plot the maximum extent of the HII region along the line of sight of the projection image as a cyan (light blue) contour. In the simulation with a magnetic field, by contrast, the smoother density field causes the HII region to become (relatively) more spherical.}

\rev{For most of the simulations in this paper we include a magnetic field. While there are major qualitative differences between the results with and without magnetic fields, the mean density field in each simulation is similar. As a result quantities such as median ionisation front radius, mass of ionised gas and momentum of the cloud gas are broadly similar. We discuss comparisons between these quantities and analytic theory in the next section.}

\section{Influence of Photon Emission Rate and Cloud Compactness}
\label{results:free}

In this section we compare our simulations to our analytic models, focussing on the influence of varying photon emission rates and cloud compactness on the properties of HII regions. In the first instance we compare our simulations to analytic models that assume a spherically symmetric power law density field with no velocity or pressure terms outside the ionisation front. The point of this exercise is to determine to what extent previous models for the expansion of ionisation fronts hold in more complex media. In subsequent sections we discuss at what point these models break down and introduce new ones that fit the simulations better. We invoke Equations \ref{overview:radius} to \ref{overview:momentum} given in Section \ref{expansion:powerlaw}.

We fit our simulated clouds to a power law density field (Equation \ref{expansion:density_profile}) using the spherically-averaged density profile sampled from the simulation output at the time the source is turned on, sampling inside a sphere of radius 13.5 pc around the source (50\% of the distance to the edge of the box). These fits are not perfect descriptions of the actual density field, but they give a reasonable agreement to the density field while allowing comparison to analytic models. It also allows a point of comparison with observed clouds, though as \cite{Tremblin2014a} note, there exists some degeneracy with regards to modelling the age and radius of observed HII regions. We do not directly address comparsion with observations in this work. In the fiducual cloud at $t_{ff}$ we find a power law index of -0.74. This gives a solution to Equation \ref{overview:radius} where $r_i \propto S_*^{0.18} t^{0.72}$.

Our simulations include a static source of UV photons in the centre of the simulation volume, turned on after one cloud free-fall time. In reality, the UV emission rate will be determined by the mass of the most massive star \citep{Vacca1996,Martins2005}, which in turn depends on the mass of the cluster formed as well as how the IMF is sampled. We will treat self-consistent star formation in future work. However, it should be noted that as the most massive stars fall towards the centre of the cloud as the cluster undergoes mass segregation, it is a reasonable approximation to assume that the majority of the UV emission rate is coming from the centre of the cloud.

In Figure \ref{flux:images}, we show the density field of the simulation for different timesteps in the simulations \simname{N00\_B02}, \simname{N47\_B02}, \simname{N48\_B02}, and \simname{N49\_B02} - that is, for simulations in the Fiducial cloud but with varying source emission rates $S_* = $ 0, $10^{47}$, $10^{48}$, and $10^{49}$ s$^{-1}$. The first and third of these simulations are repeated from Figure \ref{overview:images}. For the weakest source ($10^{47}$ s$^{-1}$), the ionisation front is insufficient to resist the infalling clumps in the right hand side of the figure. As a result the HII region expands in one direction only as a ``blister'' region, while for larger UV emission rates, the ionisation front expands in all directions. The most diffuse gas is pushed away as a dense shell, while the most massive clumps remain embedded inside the HII region. After 3 Myr in the $10^{49}$ s$^{-1}$ simulation, the cloud is nearly entirely destroyed save for a few cometary clouds, while the $10^{47}$ s$^{-1}$ source has barely changed the structure of the cloud when compared to the run with no UV source.

\subsection{Modelling the Ionisation Front Radius}
\label{results:free:radial}

We plot the properties of the HII region with varying photon emission rate in Figure \ref{flux:properties}. In Figure \ref{flux:images} we have already seen that there is a considerable scatter in the radius of the ionisation front with angle. As a result there is no single ionisation front radius as in the 1D models, but a distribution of radii. Therefore, we plot the \textit{median} radius of the ionisation front, measured by sampling the radius of the ionisation front along several lines of sight. \rev{We use the median rather than the mean as the latter biases towards extreme values, causing our results to overestimate the radius of the ionisation front. Rather, we are interested in whether the front is able to escape over the majority of lines of sight}. Comparing these results to the solution of Equation \ref{overview:radius} given the power law fit to each simulation as described above, we find a reasonable agreement between the analytic theory and the simulation results at early times in the runs with $10^{48}$ and $10^{49}$ photons/s sources. A key discrepancy between the power law density fit and the 3D cloud structure is the presence of a sharp discontinuity the at cloud edge between the dense cloud gas and the diffuse external medium. This discontinuity lies at a radius varying from 3 to 12 pc depending on the line of sight from the source. As a result the ionisation front expands rapidly into the external medium once it reaches this edge, which happens at around 1 Myr with the $10^{49}s^{-1}$ source and 1.5 Myr for the $10^{48}s^{-1}$ source. We discuss a model that corrects for this in Appendix \ref{cloudoutflow}, though for reasons of simplicity we do not use this model in our analysis.

The results for the simulation without a magnetic field (\simname{N48\_B00}) are very similar to the results with one (\simname{N48\_B02}) during the early expansion phase, since the mean density profiles are similar. However, as we discussed in Section \ref{overview:before}, the density field is more fragmented without a magnetic field. This means that the ionisation front can leave the cloud more rapidly. As a result, it accelerates away from the power law solution sooner than in the run with a magnetic field.

The radial evolution of the $10^{47}$ photons/s run in the top panel of Figure \ref{flux:properties} is significantly different from the power law model. While the ionisation front leaves the box over a small fraction of the solid angle around the source, the \textit{median} radius of the ionisation front stagnates. In other words, for at least half of the solid angle around the source, the ionisation front remains trapped by the cloud, at least for 4 Myr after the source is turned on. We return to this issue in Section \ref{results:stalled}.

\subsection{Ionised Mass and Momentum}
\label{results:free:ionmassandmom}

We can also compare the ionised mass and momentum added to the system by the HII region to the analytic expressions given in Equations \ref{overview:ionmass} and \ref{overview:momentum}. It is worth repeating that even for an ionisation front that expands to engulf the entire cloud, most of the mass in the cloud is found the neutral shell. This is because, from Equation \ref{expansion:photon_balance}, $n_i = n_0 (r_s / r_i)^{3/2}$, and hence most of the mass displaced by the ionisation front is pushed onto the dense shell around the cloud. As with the radius of the ionisation front, the mass in ionised gas for $S_* = 10^{47}$s$^{-1}$ is overestimated, since the ionisation front is trapped in the direction where most of the mass in the cloud is found. The non-zero mass in ionised gas in \simname{N48\_B00} at t=0 is because the gas in the diffuse phase is hot enough to become collisionally ionised. This does not affect our subsequent results for the photoionisation of the cloud.

For the momentum in the cloud we find a good fit between the analytic model and the simulation results until mass begins to leave the simulation volume. \rev{As in \cite{Iffrig2015} we give momentum} as the total momentum in radial flows in the simulation. The momentum in the $10^{49}$s$^{-1}$ simulation approaches the momentum found when a supernova goes off in the cloud \citep[see again][]{Iffrig2015}. The $10^{47}$s$^{-1}$ source is insufficient to displace a significant quantity of mass and the momentum of the cloud is not very different from the case without a source of photons at all. \rev{The results of the simulations with sources of 0 and $10^{47}$ photons/s are not identical due to the nonlinear nature of the flows in the cloud, but they both fluctuate between 2 and 4$\times10^{42}$ g cm/s of momentum.} The momentum in simulations \simname{N48\_B00} and \simname{N48\_B02} is very similar until the ionisation front leaves the simulation volume, which it does earlier in run \simname{N48\_B00} as explained in the previous section.

In Figure \ref{compact:properties} we plot the momentum and ionised mass in simulations with increasingly compact initial conditions. Varying the cloud compactness gives a similar effect to varying the photon emission rate. The mass in ionised gas is well captured by the Power Law model for the Fiducial and More Compact clouds, whereas in the Most Compact case we find a negligible quantity of ionised gas. Similarly, for the momentum, only in the Fiducial case does the Power Law model accurately predict the amount of momentum added to the system by the HII region. In the More Compact and Most Compact clouds, the momentum in the system is dominated by the fluctuating momentum of the cloud itself. In simulation with $10^{48}$ photons/s in the More Compact cloud, there is an initial divergence in the results but this only maintains the momentum in radial flows, compared to a predicted drop in momentum in radial flows when there is no source of UV photons. It should be noted that the Power Law model does not predict a large increase in momentum in the Most Compact cloud, and that the model lies within the variation in momentum over time in this cloud. We return to reasons why the HII region is ineffective at driving flows in compact clouds in the following section.

\subsection{Delayed UV Emission}
\label{results:free:delayed}

One additional test we perform is to determine what happens when we allow the cloud to relax over a longer period of time before we turn the source on. To do this we use the Fiducial cloud with a source of $S_* =10^{48}$s$^{-1}$. In addition to \simname{N49\_B02} we run a further two simulations in which we turn on the source at 2 and 3 $t_{ff}$ (2.5 and 3.75 Myr), \simname{N48\_B02\_F2} and \simname{N48\_B02\_F3} respectively. At $2~t_{ff}$ the power law index of the cloud density profile becomes -0.87, and at $3~t_{ff}$ it becomes -1.88 (compared to -0.74 at $t_{ff}$). Nonetheless, the resulting expansion of the HII region does not significantly change between these simulations. There is a short period during which the ionisation front stalls in both delayed simulations, but after 1 Myr the front breaks out and expands similarly to \simname{N48\_B02}. This suggests that the properties of a HII region are not very sensitive to when stars are formed in a cloud with a given initial structure, though there may be a short period during which the front must escape the increased density in the centre of the cloud.

 \begin{figure*}
 \centerline{\includegraphics[width=0.98\hsize]{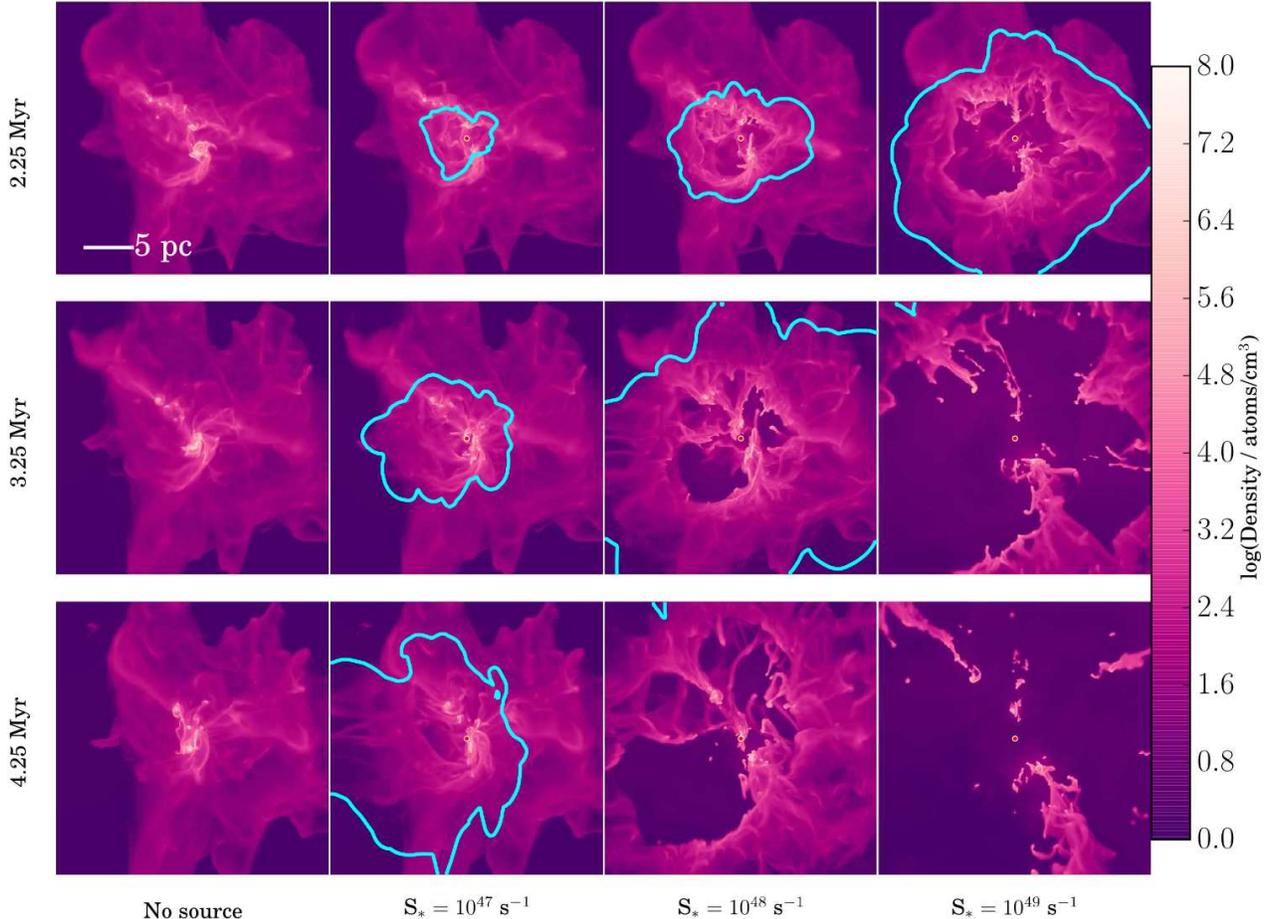}}
 \caption{As in Figure \ref{overview:images} but for varying UV photon fluxes. From left to right the columns are for simulations \simname{N00\_B02},  \simname{N47\_B02}, \simname{N48\_B02} and \simname{N49\_B02}. The rows show each simulation at $t_{ff}$ + [0,1,2,3] Myr respectively. The cloud at $t_{ff}$ is shown in the top left panel in Figure \protect\ref{overview:images}.}
  \label{flux:images}
 \end{figure*}

  \begin{figure}
 \centerline{\includegraphics[width=0.98\hsize]{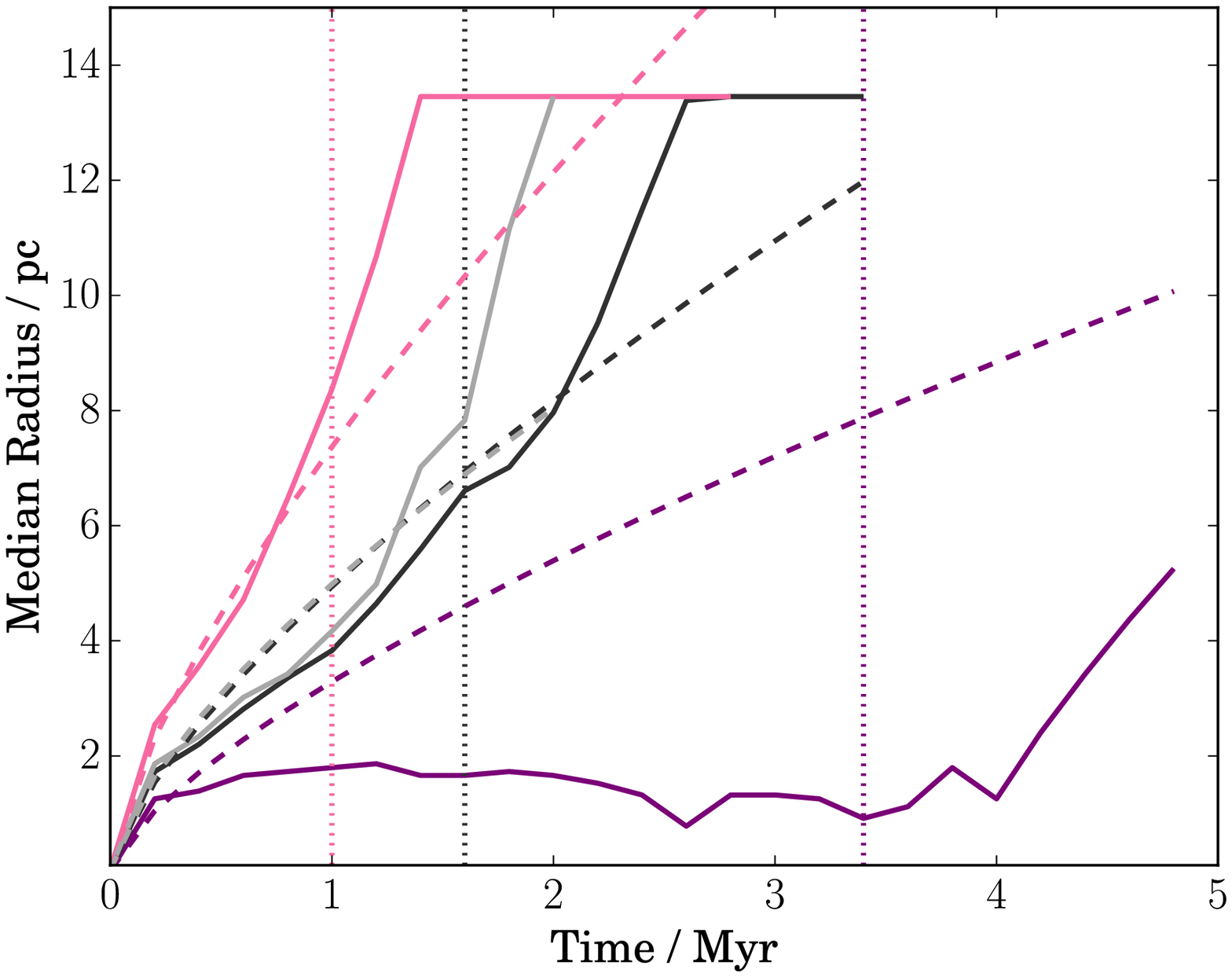}}
 \centerline{\includegraphics[width=0.98\hsize]{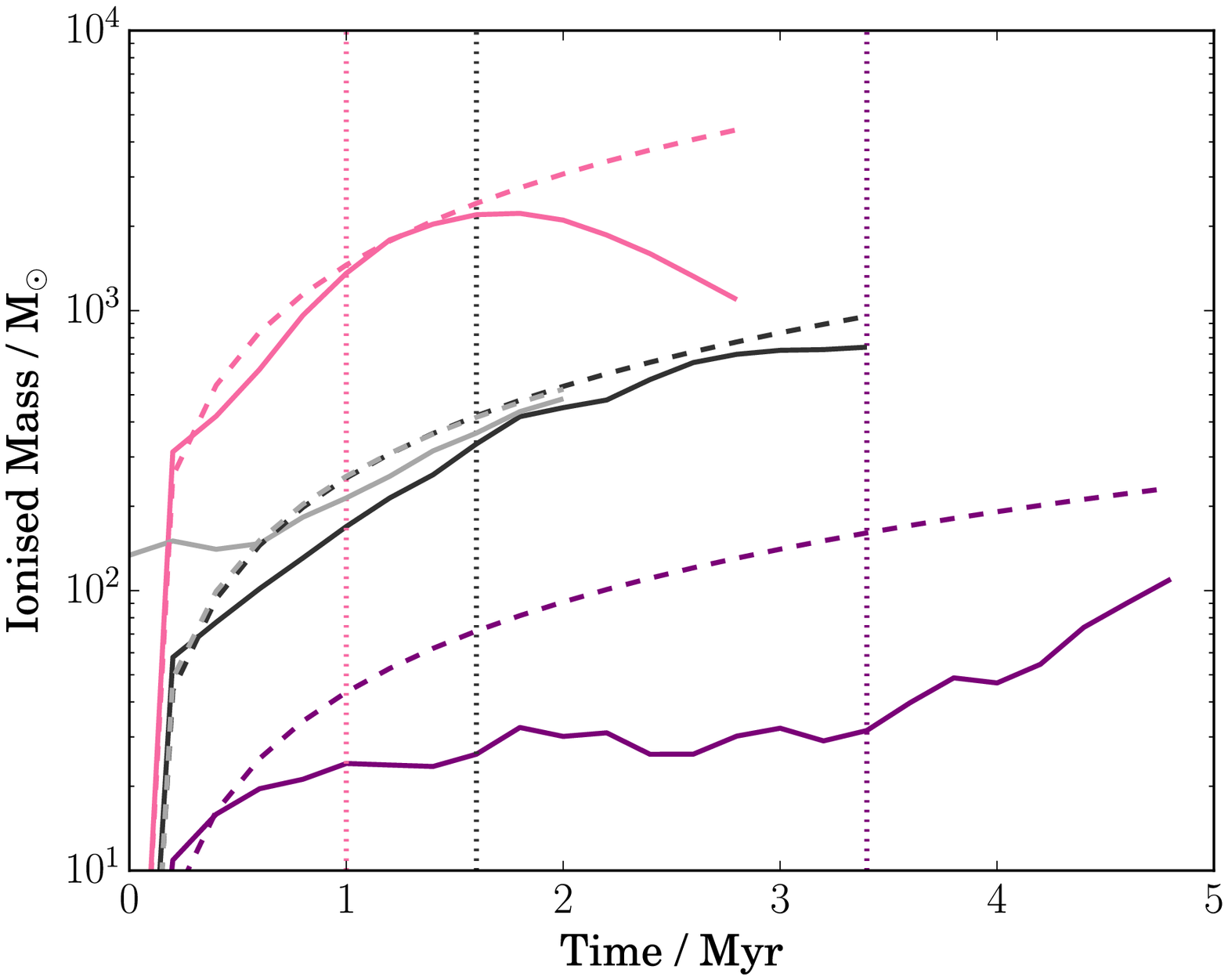}}
 \centerline{\includegraphics[width=0.98\hsize]{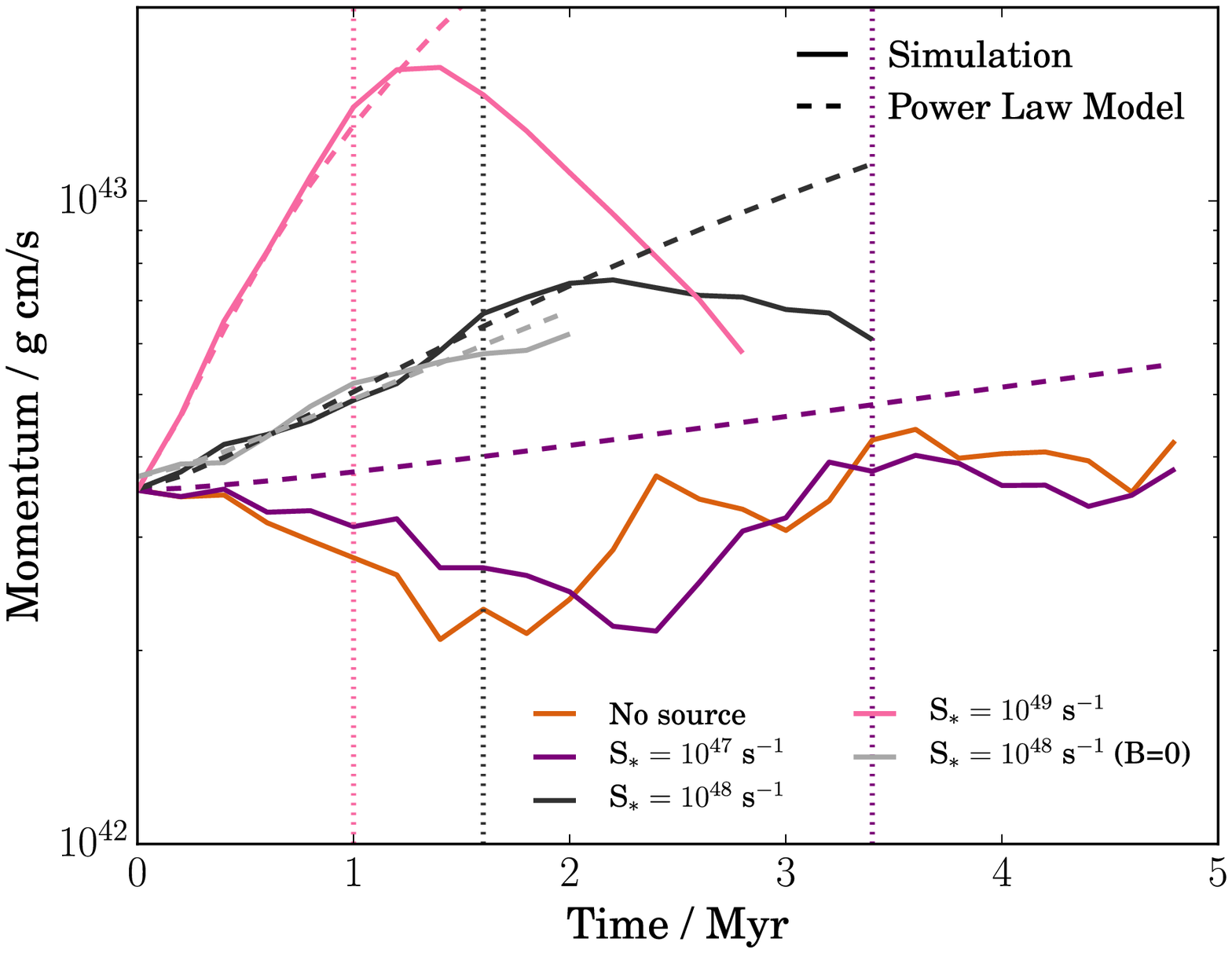}}
 \caption{Properties of the HII region over time for varying UV photon emission rates, with the same simulations as in Figure \ref{flux:images} plus \simname{N48\_B00}. In the top panel is the median radius of the ionisation front across randomly sampled lines of sight. In the middle panel is the mass in ionised gas. On the bottom is the momentum in radial flows in the simulation. \rev{Each colour corresponds to a different photon emission rate. The simulation results are shown as a solid line. We plot as a dashed line the Power Law model (see Section \ref{expansion:powerlaw}) for each photon emission rate.} The time given is from the time the source is turned on ($t_{ff}=1.25~$Myr after the start of the simulation). We overplot as dotted lines the time in each simulation where the ionisation front first leaves the simulation volume.}
  \label{flux:properties}
 \end{figure}

  \begin{figure}
 \centerline{\includegraphics[width=0.98\hsize]{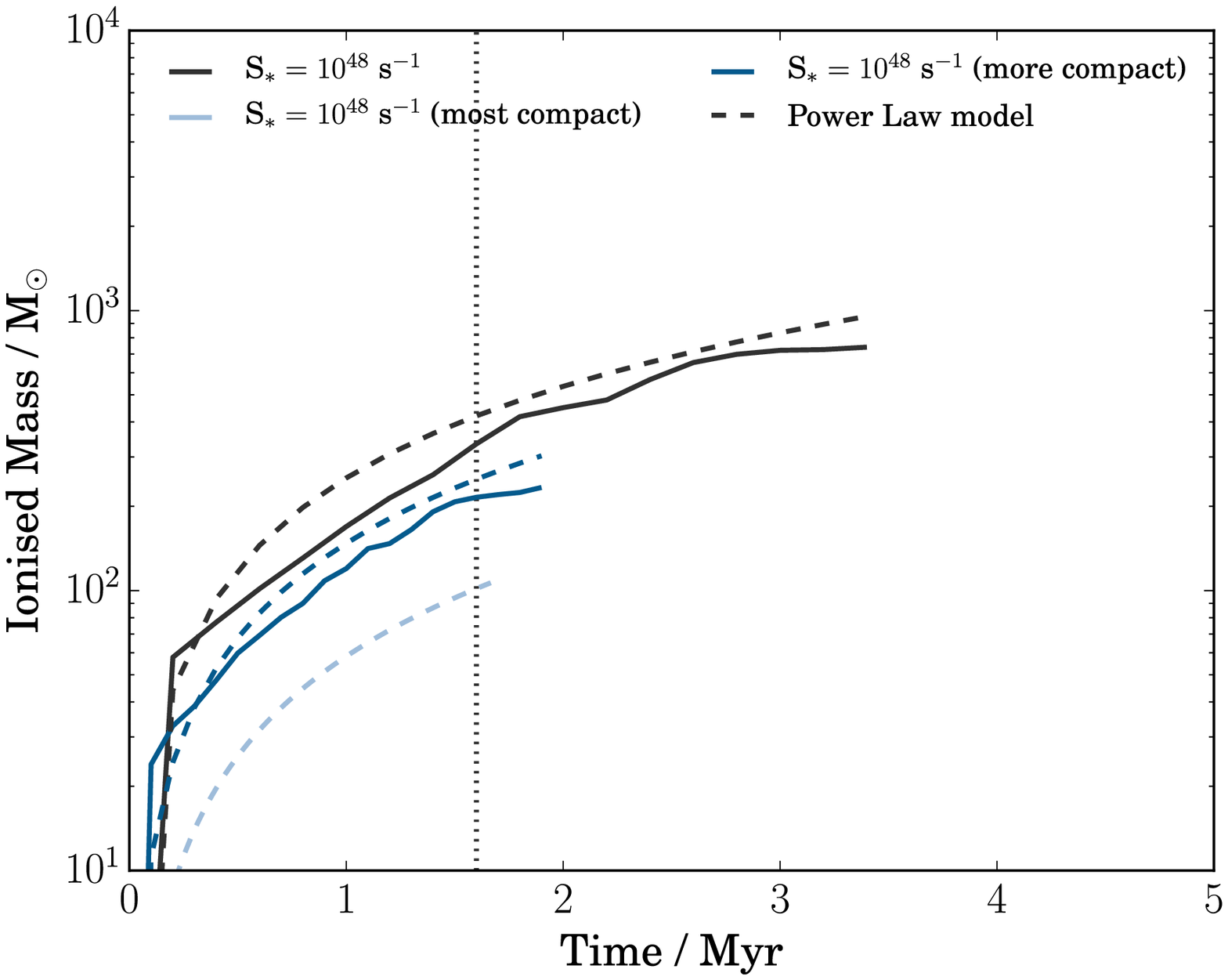}}
 \centerline{\includegraphics[width=0.98\hsize]{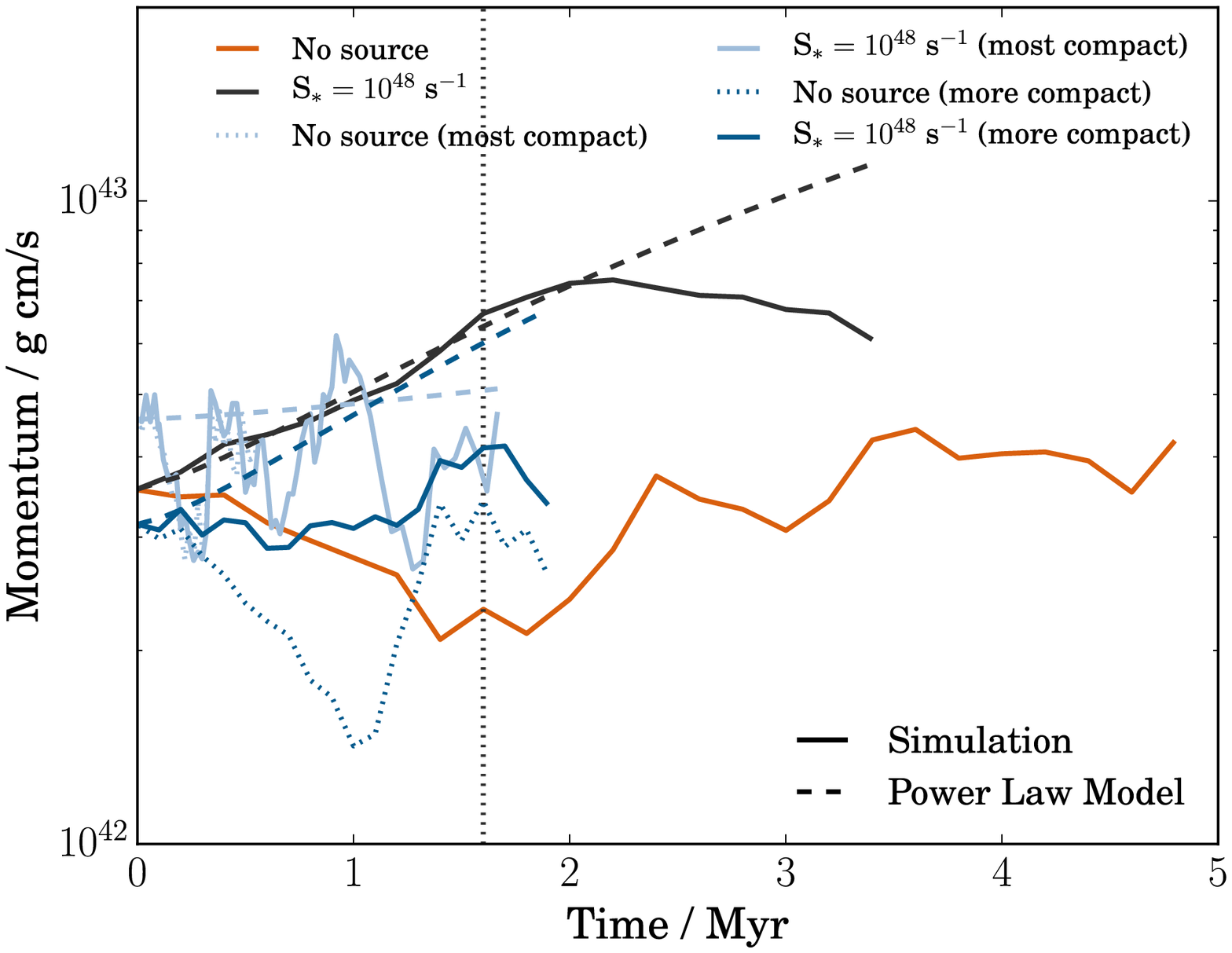}}
 \caption{Properties of the HII region over time for varying cloud compactness, as in Figure \protect\ref{flux:properties}. \rev{Each colour corresponds to a different cloud ``compactness''. The solid lines show the simulation results for each cloud compactness with a photon source. The dotted lines show the  simulation results for the same cloud without a photon source. The dashed lines show the Power Law model (see Section \ref{expansion:powerlaw}) for each cloud compactness.} The time axis begins at $t_{ff}$ after the start of the simulation, for $t_{ff}$ as defined in Table \protect\ref{methods:numsimtable}.}
  \label{compact:properties}
 \end{figure}
 
%   \begin{figure}
%  \centerline{\includegraphics[width=0.98\hsize]{plots/tstart/medianradius.eps}}
%  \caption{As for the first plot showing median radius in Figure \ref{flux:properties} but for the delayed source comparison simulations \simname{N48\_B02}, \simname{N48\_B02\_F2} and \simname{N48\_B02\_F3}.}
%   \label{limits:tstart}
%  \end{figure}

\section{Stalled Expansion}
\label{results:stalled}

In the previous section we compared our models assuming a static power law density profile to our simulations in the Fiducial cloud. These models provided a good fit to the simulations provided the UV photon emission rate was above $10^{48}$ photons/s. However, the front around the $10^{47}$ photons/s source stalled and was unable to expand further for 4 Myr. In this section we compare our model for expansion in a medium with a non-zero external velocity field given in Section \ref{expansion:raga} in order to explain this effect. 

\rev{In Figures \ref{flux:properties} and \ref{compact:properties} we plotted the Power Law model assuming no external pressure terms. This model fit the simulation results well for large photon emission rates or less compact clouds, but not so well for weak sources or dense clouds. In this section we address this by solving Equation \ref{expansion:raga_like} including a velocity field. We plot this solution as a ``Non-Static Model'' (as opposed to the model in the previous section that assumed a static cloud with no velocity field) next to our simulation results in Figure \ref{stalled:ragacompare}.}

\rev{In each solution to this equation use two approximations for the density and velocity field}. In the ``Power Law'' model, we assume a power law density profile as in the previous section, with a velocity profile calculated as the escape velocity at each radius as in Equation \ref{virialised:vesc}. This model is highly idealised, and so we also plot the ``Sampled'' model, which samples the time-dependent density and velocity field from an identical simulation without a source of photons. The values sampled are taken from the spherically-averaged radial density and velocity profiles in each output and interpolated in radius and time. The ``Power Law'' model provides a close comparison with our simple analytic model, whereas the ``Sampled'' model gives a closer match with our simulated cloud. Both models however assume a spherical density and velocity field. We discuss the limitations of this assumption below.

\subsection{Free vs Stalled Expansion}
\label{results:stalled:freevsstalled}

\begin{figure*}
\centerline{\includegraphics[width=0.48\hsize]{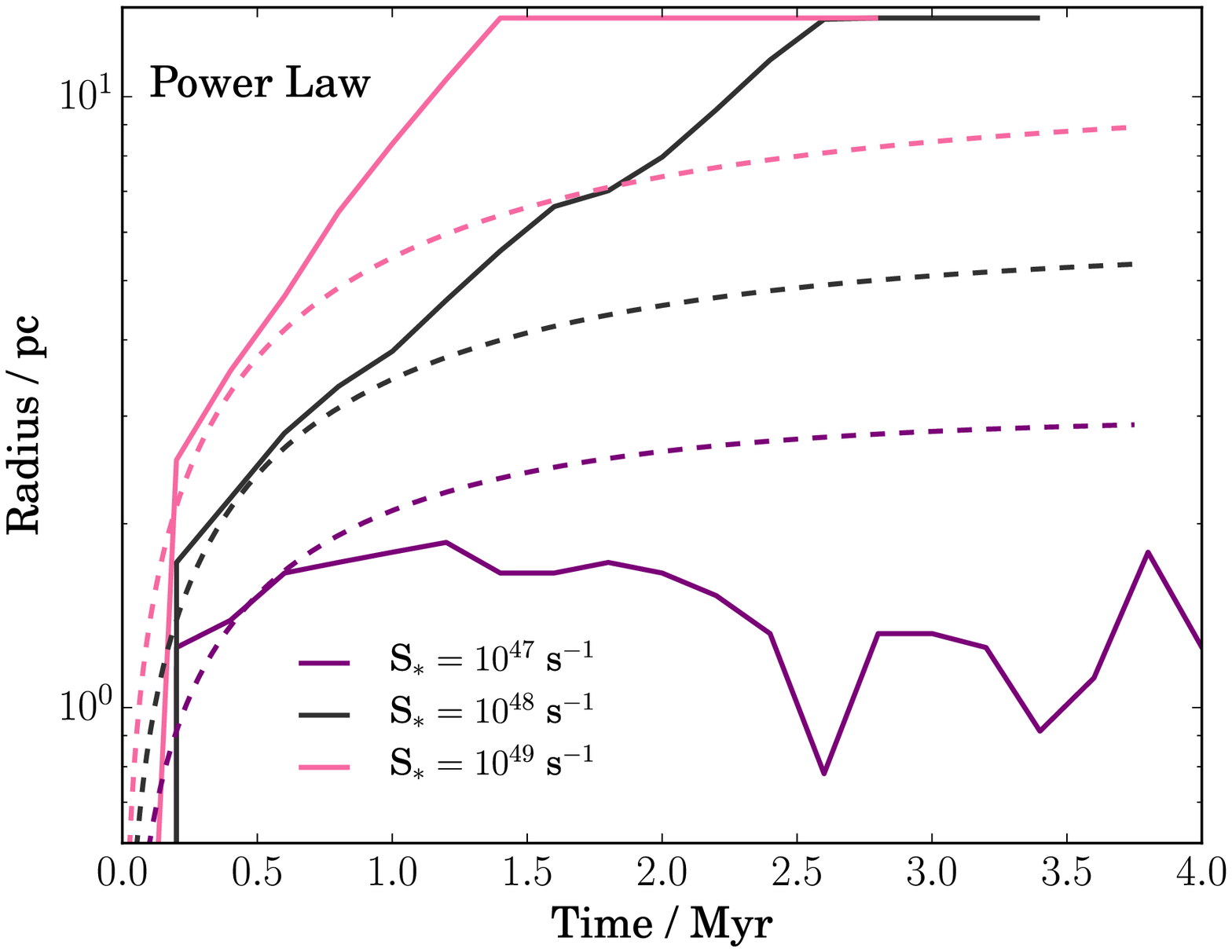} \includegraphics[width=0.48\hsize]{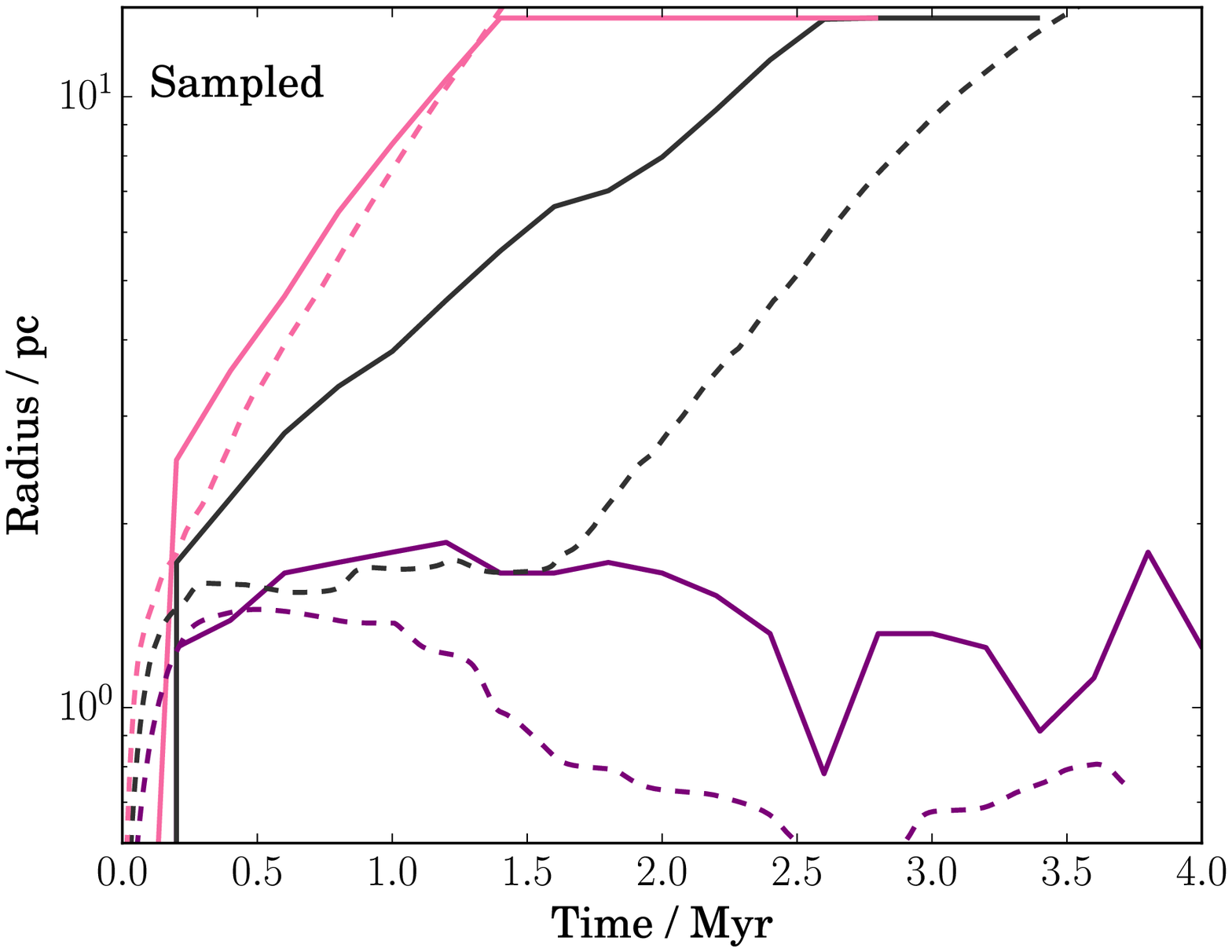}}
\centerline{\includegraphics[width=0.48\hsize]{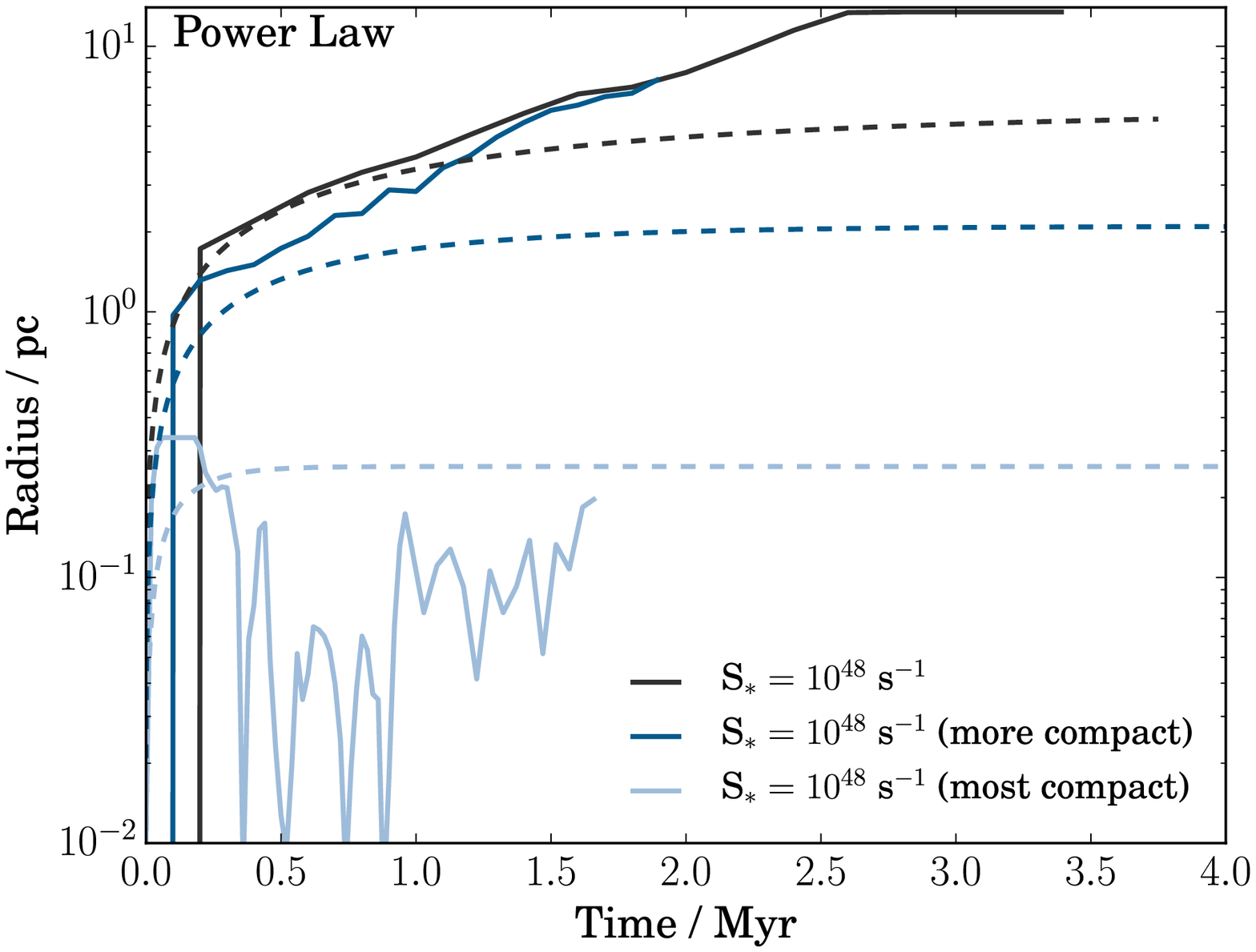} \includegraphics[width=0.48\hsize]{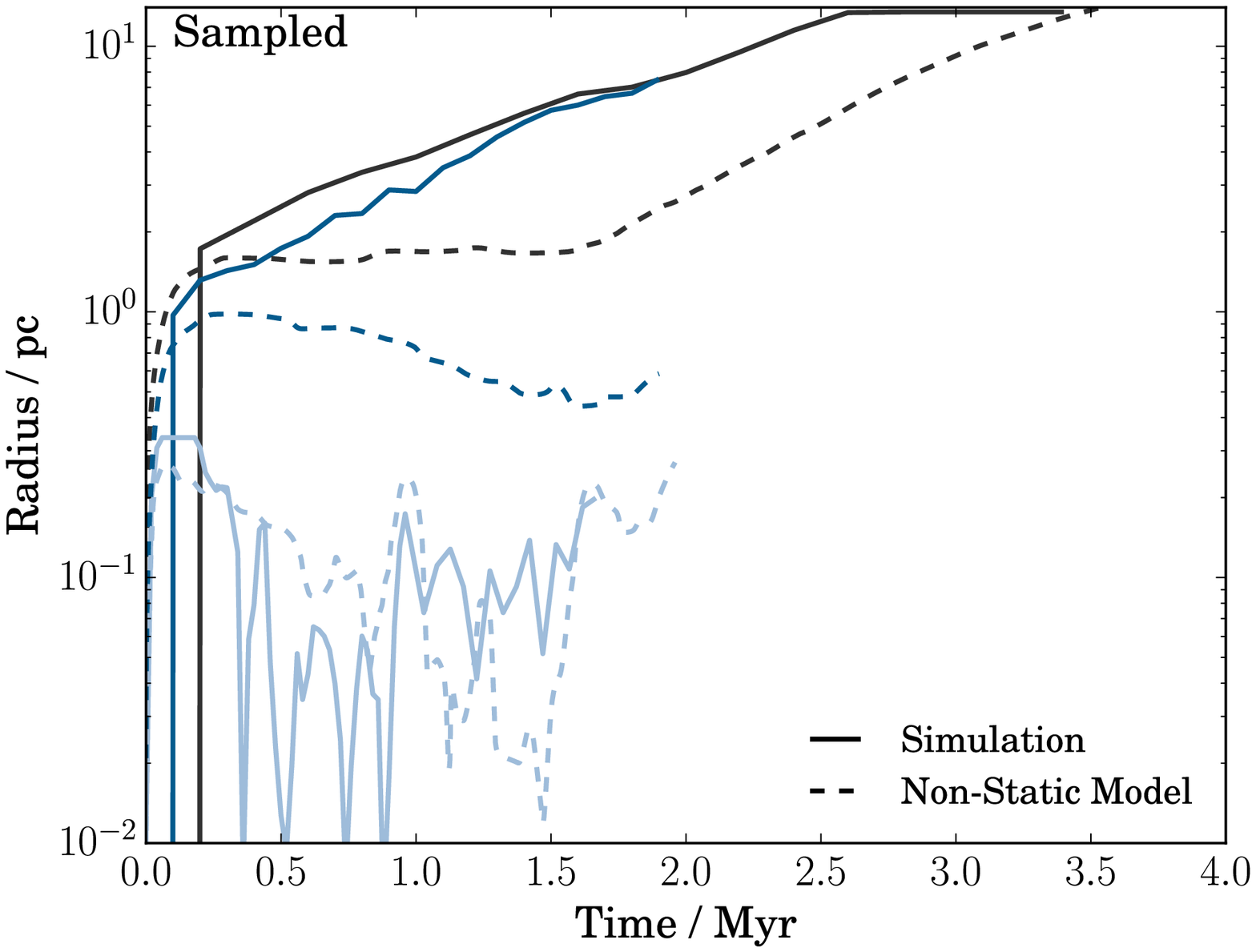}}
\caption{The median radius of the ionisation front over time compared to Equation \protect\ref{expansion:raga_like}. The top panels show simulations \simname{N47\_B02},  \simname{N48\_B02} and  \simname{N49\_B02}. The bottom panels show simulations \simname{N48\_B02}, \simname{N48\_B02\_C} and \simname{N48\_B02\_C2}. The left panel uses a model assuming a power law density profile and velocity profile given by the escape velocity at each radius $\sqrt{2GM(<r)/r}$. The right panel models the density and velocity field of each simulations directly by sampling the density and velocity at each radius and time in identical simulation without a source of photons. The solid lines show the median radius of the ionisation front in each simulation. The dashed lines show the \rev{``Non-Static Model'', in which we solve Equation  \protect\ref{expansion:raga_like} including} an external velocity field.}
 \label{stalled:ragacompare}
\end{figure*}

In the top panels of Figure \ref{stalled:ragacompare} we use the Fiducial cloud while varying the photon emission rate. The $10^{48}$ and $10^{49}$ photons/s sources cause the ionisation front to break out of the cloud, while the front stalls with a $10^{47}$ photon/s source. Similarly, in the bottom panels, the More Compact and Fiducial cloud allow the ionisation front to expand freely while in the Most Compact cloud the front stalls. \rev{Our solution to Equation \ref{expansion:raga_like} including a velocity field thus matches our simulation results better for the weak source and most compact cloud.}

The Power Law and Sampled models both predict the radius at which the ionisation front stalls in the simulations where this occurs. The Sampled model also reproduces the expansion and contraction of the ionisation front in the case where it stalls. We discuss the cause of this in Section \ref{results:stalled:clumps}. The Sampled model predicts incorrectly that \simname{N48\_B02} should stall for around 2 Myr, since it assumes that that the density field is spherically symmetric, meaning that the ionisation front cannot escape through channels of low density as it does in the simulation.

The Power Law model tends towards the value for $r_{stall}$ given in Equation \ref{virialised:rstall}. In this equation, $r_{stall}$ varies as $(S_*^{1/4} / n_0)^{\psi}$. $\psi$ is larger for steeper density profiles. This means that the effect of cloud density and photon emission rate is slightly enhanced in the Most Compact cloud, which has $w \simeq 1$ ($\psi = 0.8$), compared to the Fiducial cloud, which has $w \simeq 3/4$ ($\psi = 0.72$). We give the value for $r_{stall}$ in each simulation in Table \ref{freevsstalled:rstalltable}. In none of our simulations do we find that the escape velocity exceeds the sound speed of the gas, and hence in all simulations $r_{stall} > r_s$.

In the following subsections we discuss the reason why \rev{the ionisation fronts in some simulations seem to expand freely into the external medium, while in others it stalls. We also discuss} why $r_{stall}$ in the simulations and in the Sampled model appears to oscillate \rev{(causing the HII region to ``flicker'')}.

\begin{table}
\begin{tabular}{l c c c c c c c c c}
   \textbf{Simulation} & \textbf{$r_{cloud}$/pc} & \textbf{$r_{stall}$/pc} & \textbf{$r_{stall} / r_{cloud}$} \\
  \hline
 \simname{N47\_B02} & 3.0 & 3.55 & 1.18  \\
 \simname{N48\_B02} & 3.0 & 6.24 & 2.08 \\
 \simname{N49\_B02} & 3.0 & 10.4 & 3.47  \\
 \simname{N48\_B02\_C2} & 1.7 & 2.83 & 1.68  \\
 \simname{N48\_B02\_C} & 1.0 & 0.433 & 0.577 \\
  \hline
\end{tabular}
  \caption{Table comparing the minimum cloud radius $r_{cloud}$ (as measured at $t_{ff}$ in each simulation) to the radius at which the ionisation front stalls in the analytic model $r_{stall}$. If the cloud radius is smaller than the stalling radius, the front can escape the cloud.}
\label{freevsstalled:rstalltable} 
\end{table}

\subsection{Role of the Diffuse Medium}
\label{results:stalled:diffuse}

The assumption of spherical symmetry in the density and velocity field breaks down when we reach the cloud edge at $r_{cloud}$. The cloud edge is a sharp transition of a fraction of a parsec between the dense cloud gas and the diffuse external medium. The Power Law model does not resolve this edge because $r_{cloud}$ varies over a large range (3 to 12 pc in the Fiducial cloud). If $r_{stall} / r_{cloud} > 1$, the front will enter the diffuse medium, where the ram pressure becomes very low since the density and velocity are much lower than inside the cloud. Hence beyond this radius the front will no longer stall, but expand very rapidly outwards.

In Table \ref{freevsstalled:rstalltable} we give the values for $r_{cloud}$, $r_{stall}$ and $r_{stall} / r_{cloud}$. The latter value is close to 1 in simulation \simname{N47\_B02}. This means that (on average) the front is only just expected to escape the cloud. If we compare $r_{stall}$ in this simulation to Figure \ref{stalled:ragacompare}, we find that it has not yet reached $r_{stall}$. However, in Figure \ref{flux:properties}, we do see a sudden expansion of the mean ionisation front radius at 4 Myr, which can be attributed to the ionisation front breaking out of the cloud. In simulation \simname{N48\_B02\_C}, the ratio is 0.577. Thus the ionisation front is expected to remain trapped inside the cloud. In all other simulations $r_{stall}$ is much larger than $r_{cloud}$. This means that the stalling criterion does not affect the ionisation front beyond this radius and expand freely into the diffuse medium. 

We thus have a criterion to determine whether the median radius of the ionisation front should follow the static Power Law model or a model that includes ram pressure from the external velocity field. This criterion is very important in determining the other properties of the HII region and cloud, such as momentum injected into the \ISM as well as the strength of star formation feedback. In the next section we discuss the role that density inhomogeneities play in regulating the expansion of the HII region.

\subsection{Role of Dense Clumps}
\label{results:stalled:clumps}

Our simulations (and observed molecular clouds) are highly non-spherical, with filamentary and clumpy structures. While our spherically symmetric 1D analytic models work well despite this, there are some effects caused by these inhomogeneities that we now discuss.

Dense clumps can resist the expansion of the HII region due to having a higher density than the rest of the cloud and a smaller surface area. The most massive clumps remain embedded inside the HII region as cometary clouds. They block the expansion of the HII region over the solid angle subtended by the clump from the source, creating a ``shadow'' of neutral gas behind them. If this angle is small enough, it does not affect the median radius. However, if they move close to the source they can increase the density into which the front expands ($n_{ext}$), especially if they pass through the position of the source. This prevents the ionisation front from ionising and expelling a larger fraction of the cloud.

\begin{figure*}
\centerline{\includegraphics[width=0.48\hsize]{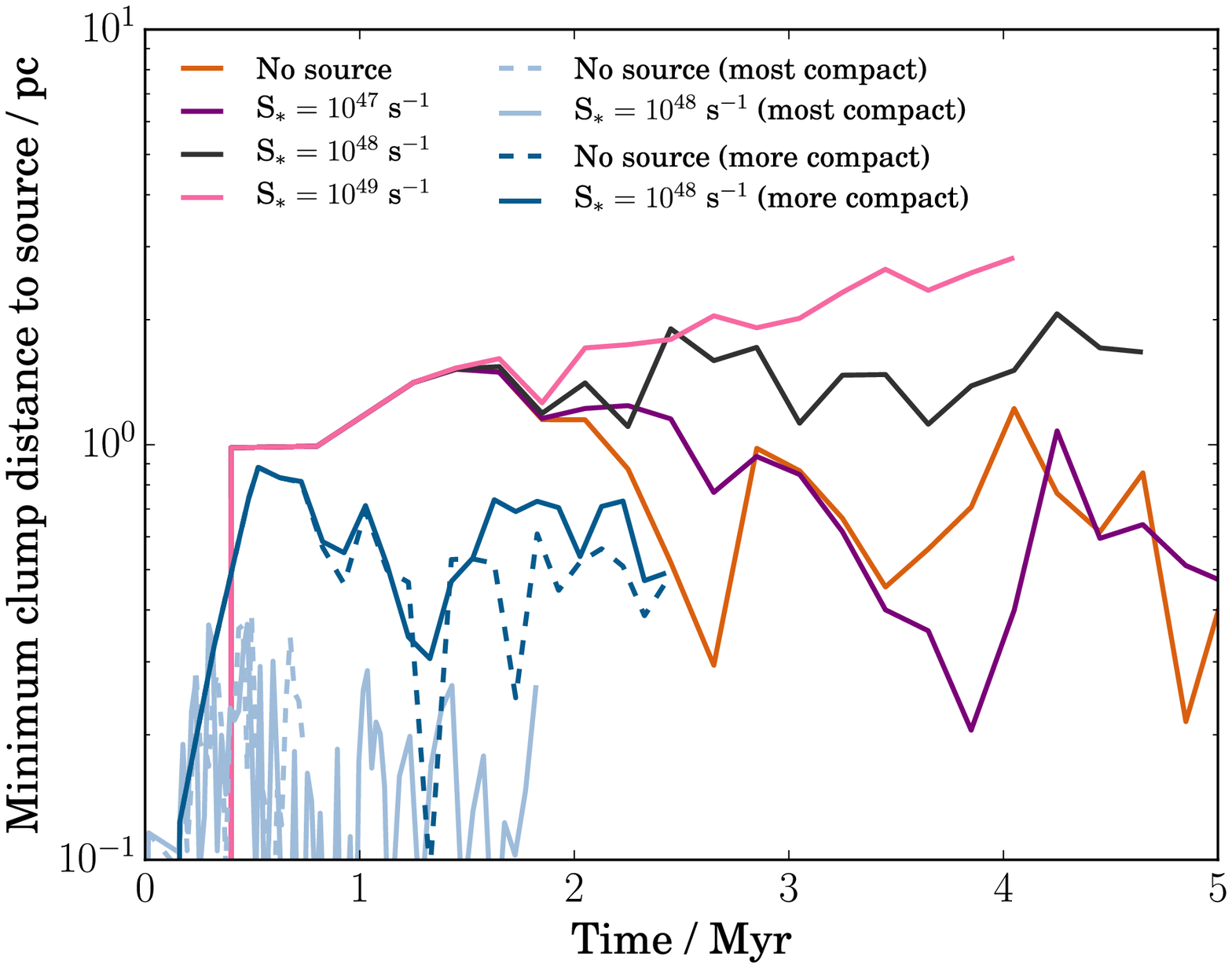}
\includegraphics[width=0.48\hsize]{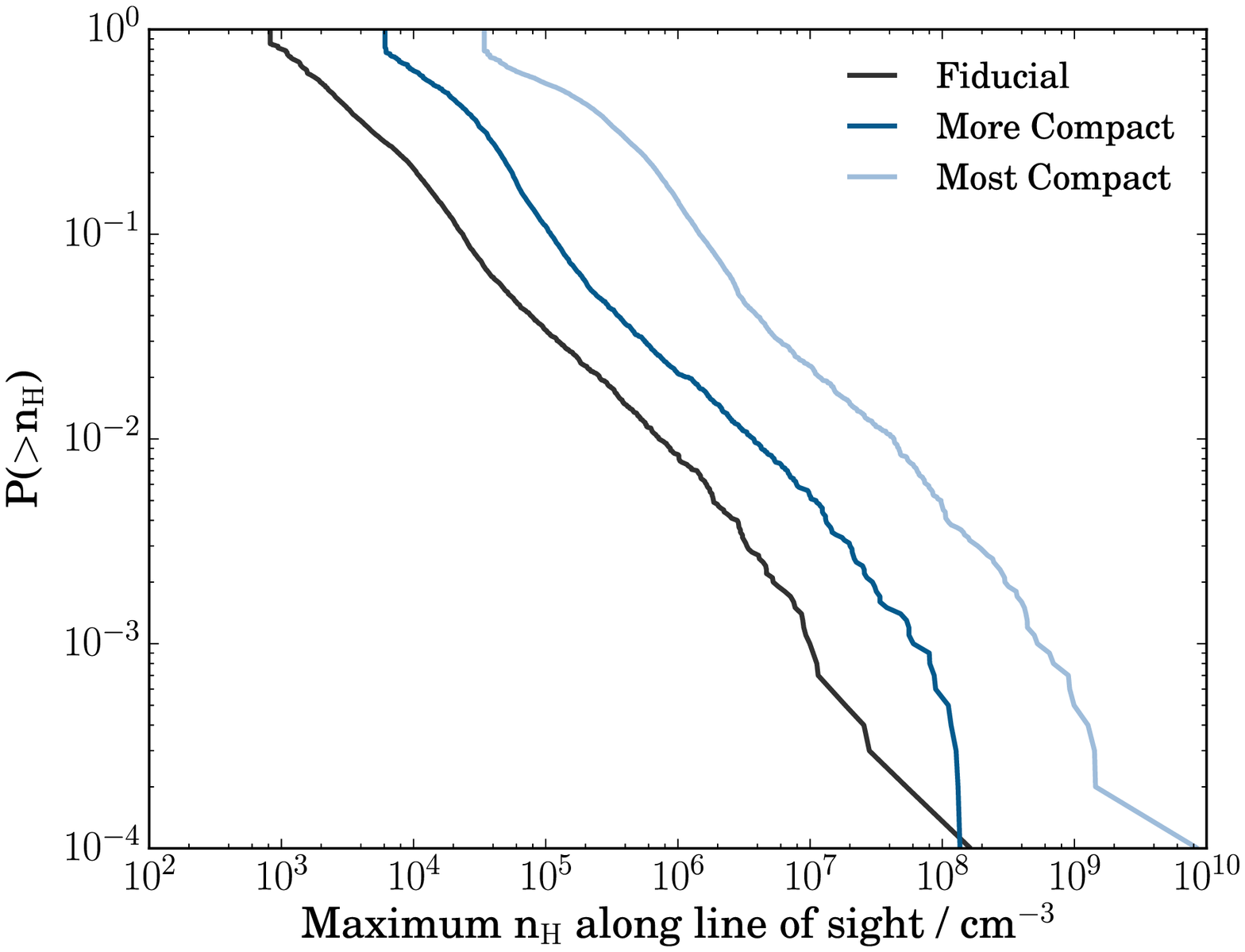}}
\caption{Left: Minimum distance from the edge of the nearest dense gas clump to the source position in simulations over time. Right: the probability that a ray cast from the source in a random direction at $t_{ff}$ in each simulation will encounter material of at least a given density in each of the clouds of a given compactness.}
 \label{freevsstalled:clumps}
\end{figure*}

In Figure \ref{freevsstalled:clumps} we plot the time evolution of the distance between the source and the edge of the nearest clump. We define this as the smallest distance between the source and a cell above a given density threshold. This threshold is $10^6$ \atcc in the Fiducial cloud, $10^7$ \atcc in the More Compact cloud, and $10^8$ \atcc in the most compact cloud. This is because the cell sizes in the denser simulations are smaller, allowing us to resolve gas at higher densities. The thresholds for defining which gas is in dense clumps are found such that there is a 1\% chance that a ray cast from the source to the edge of the cloud at $t_{ff}$ in a random direction will encounter a cell of at least this density. This is seen in the right hand plot of the same figure, where we plot the probability distribution function of maximum density along the lines of sight from the source position to the edge of the simulation volume. The reason the ionisation front is non-spherical is due to this probability distribution function of densities around the source.

As we increase the photon emission rate, the clumps are accelerated further from the source, meaning that they play little role in the expansion of the ionisation front in other directions. As such, for higher photon fluxes in the Fiducial cloud, the front can expand freely in most directions. The $10^{47}$ photons/s source is insufficient to push away the clumps. The radius at which the front stalls is thus affected in part by the motion of the clumps. The radius of the ionisation front oscillates in Figure \ref{stalled:ragacompare} as the clumps orbit close to the source.

In the denser clouds, the source is largely unsuccessful at significantly altering the trajectory of the dense clumps. In the More Compact cloud, the clumps remain far enough from the source for the ionisation front to escape over at least half of the solid angle around the source. By contrast, in the Most Compact cloud the clumps orbit close to the source, passing through it multiple times. As a result the source flickers on and off, as seen in the simulations results and the ``Non-Static'' model in Figure \ref{stalled:ragacompare}.

The motion of dense clumps inside the cloud thus plays a role in determining the shape and behaviour of the HII region. However, the broad behaviour of the HII region can be described well with a spherically symmetric model that takes into account the distance between the source and the edge of the cloud.

\section{Evaporation of Star-Forming Clumps}
\label{results:feedback}

Molecular clouds are the sites of star formation in galaxies. An important role of UV photoionisation is thus the self-regulation of star formation. This introduces feedback cycles from stars, which both photoionise the clouds in which they formed and, if the ionisation fronts escape the cloud, affect the evolution of nearby clouds. This feedback can either be positive, as in the case of shock compression enhancing the star formation rate, or negative, as in the case of UV photons evaporating clouds that would otherwise be star forming. In this work we do not simulate star formation directly, nor do we place sources of UV photons onto star particles. We thus do not reproduce these feedback cycles directly in our simulations. Instead we discuss the potential for star formation in our simulations based on stability criteria, and compare this to analytic models from the literature.

\subsection{Jeans Unstable Mass}
\label{feedback:jeansunstable}

Our highest resolution in the Fiducial cloud is 0.026 pc (= 5500 AU), which is below the radius of star-forming cores \citep{Ward-Thompson1994} but insufficient to follow their collapse into protostellar cores. We hence argue that any gas cell that is Jeans unstable is to be considered potentially ``star-forming''. The Jeans stability criterion states that if the free-fall time of any given part of the cloud is smaller than the sound-crossing time, the cloud is vulnerable to fragmentation and collapse. We set the Jeans Length $\lambda_J$ to our cell length and count the mass in cells for which
\begin{equation}
 \frac{c_s}{\lambda_J\sqrt{G \rho}} < 1
 \label{feedback:jeans_instability}
\end{equation} where $c_s$ is the sound speed in the cell and $\rho$ is the density. For example, a gas cell at our maximum resolution in the Fiducial cloud with temperature 50 K and $2 \times 10^6$ \atcc is at the Jeans stability limit. In the More Compact and Most Compact clouds the size of a cell at the maximum spatial resolution is reduced by a factor $0.75^2$ and $0.5^2$ respectively, proportionally to the radius of the cloud. It is important to note that not all mass that fulfills the Jeans criterion will end up in stars. \cite{Matzner2000,Alves2007} suggest that the star-formation efficiency (SFE) of cores is around 30\%. The mass in Jeans-unstable gas is thus an overestimate by a factor of 3-4.

We plot the total mass in Jeans unstable gas in Figure \ref{feedback:sfmass_photons}. In the Fiducial cloud with varying photon emission rate, the amount of potentially star-forming mass drops with increasing photon emission rate, even with the $10^{47}$ s$^{-1}$ source, whose median ionisation front radius is trapped by the cloud. Mass in dense clumps is evaporated and redistributed to the shell around the ionisation front. We do not find that this shell is unstable to fragmentation (see Appendix \ref{feedback:shell}). In cases where the ionisation front leaves the simulation volume, a certain amount of mass disappears from the simulation, as in the $10^{49}$ s$^{-1}$ source. However, many of the most massive clumps remain embedded inside the HII region (see Figure \ref{flux:images}). 

In the More Compact and Most Compact clouds, the radiation has little effect on the Jeans unstable mass. Even in the case of the More Compact cloud, where the front escapes the cloud, the densest gas in the simulation remains largely unaffected by the UV photons. In the previous section we explained how, despite the fact that the ionisation front can escape the More Compact cloud, the trajectories of dense clumps in the cloud are largely unaffected by the HII region.

\subsection{Modelling Clump Evaporation}
\label{feedback:clumps}

In this subsection we compare our simulation results to \cite{Bertoldi1990,Lefloch1994,Johnstone1998,Whitworth2004}, who give analytic models for the evaporation of dense gas clouds by sources of UV radiation. In our simulations we do not track individual clumps of gas. This requires special treatment in Eulerian codes such as \simname{RAMSES}, which do not trace fluid parcels as Lagrangian codes do. Instead, we sample the dense clumps in our simulation in one snapshot at $t_{ff}$ and model their mass evolution using the equations of \cite{Bertoldi1990}. \rev{We define our dense clumps to be those portions of gas in the cloud that are Jeans unstable. \cite{Bertoldi1990}} give, for thermally supported clumps, a clumps mass with a fraction \begin{equation}
(1 - t / t_{ev})^{5/3}
\label{feedback:bertoldi_massloss}
\end{equation} of its mass at $t=0$. $t_{ev}$, the characteristic evaporation time of the clump, is given by $0.448 \theta_1 c_5^{-6/5} S_{49}^{-1/5} R_1^{2/5} m_1^{2/5}~$Myr, where $\theta_1$ is a factor close to unity, $c_5$ is the sound speed of the clump in km/s, $S_{49} \equiv S_* / 10^{49}$, $R_1$ is the distance of the clump to the source in pc and $m_1$ is the initial mass of the clump in \Msolar. For the clumps in our simulation, $t_{ev}$ varies from between 1 to 30 Myr. 

We sample the clumps in our simulations at $t_{ff}$ using the following clump finding algorithm. We first select all gas cells above the Jeans stability threshold as given previously. We then assign all contiguous cells above this threshold to a single clump. For each clump we calculate the total mass of the cells inside it, its radius (measured as the maximum distance from the densest cell in the clump to any other cell) and the distance between the densest cell and the photon source. We find that at $t_{ff}$ the clump mass $m_1$, radius $r_1$ and distance to source $R_1$ are related as $m_1 \propto r_1^{5/2}$ (i.e. a Larson relation) and $m_1 \propto R_1^{-5/2}$ (measured from our simulation), though there is a large degree of scatter in the latter relationship. This is in agreement with \cite{Bertoldi1990}.

\begin{figure*}
\centerline{\includegraphics[width=0.48\hsize]{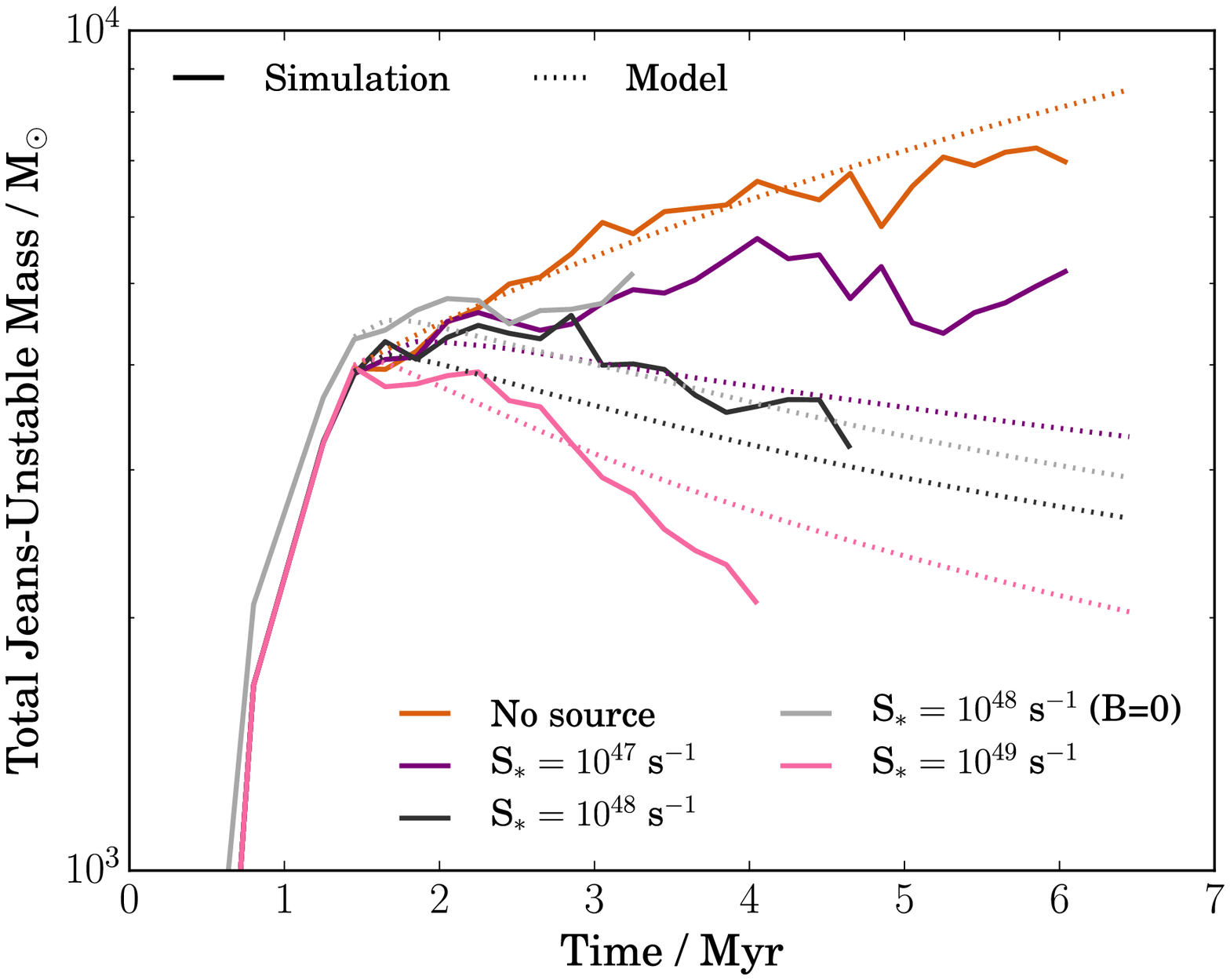}
\includegraphics[width=0.48\hsize]{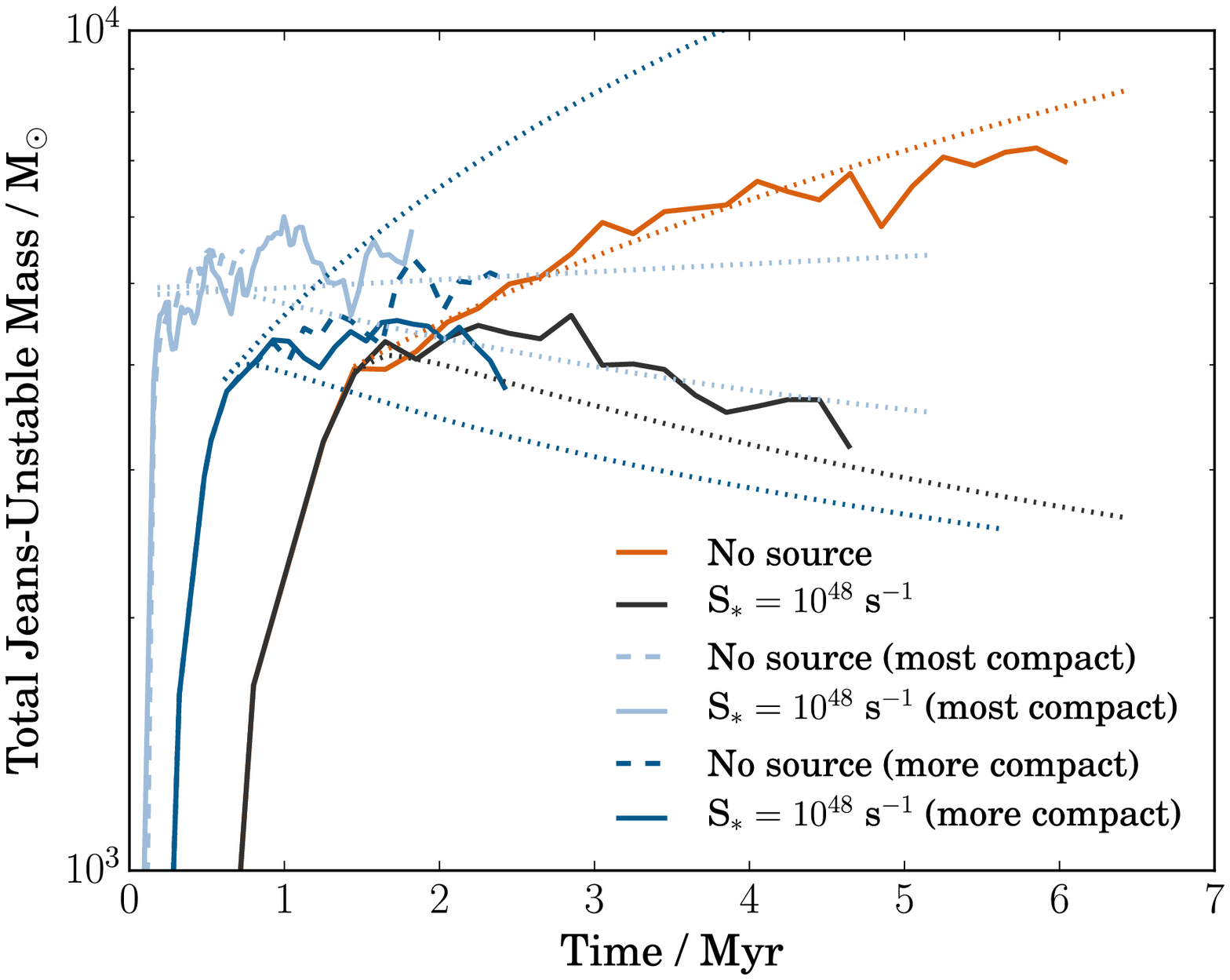}}
\caption{The total mass of Jeans-unstable gas over time for each simulation, starting from the initial conditions. The left-hand panel shows simulations in the case where the photon emission rate is changed, \simname{N00\_B02}, \simname{N47\_B02}, \simname{N48\_B00}, \simname{N48\_B02} and \simname{N49\_B02}. The right-hand panel shows simulations in which the cloud compactness is changed, \simname{N00\_B02}, \simname{N48\_B02}, \simname{N48\_B02\_C2} and \simname{N48\_B02\_C}. Overplotted as dotted lines is the model for each UV photon emission rate as described in Section \ref{results:feedback}.}
 \label{feedback:sfmass_photons}
\end{figure*}

From $t_{ff}$, we allow the mass of each clump to vary according to the following model. First, we assume that the mass of the clumps increase due to accretion as $m(t) = m_1 (1 + t/t_{acc})$, where $t_{acc}$ is set to 4.7 Myr (the free-fall time for a density of 93 \atcc, a sphere of which surrounds our cloud in the initial conditions). We omit this accretion in the Most Compact cloud since there is almost no gas in this phase in our initial conditions. Secondly, as the ionisation front reaches the clump (using the \rev{Power Law model for the expansion of the ionisation front}) we turn off the accretion and begin evaporating the clump according to Equation \ref{feedback:bertoldi_massloss}. We do not attempt to follow the orbits of the clumps as this requires more detailed modelling of the cloud dynamics that is beyond the scope of this paper. Instead, we make the simplifying assumption that the clumps remain at the same distance from the source at all times. We then sum the masses of each clump at each time and overplot as a dotted line in Figure \ref{feedback:sfmass_photons}.

For the Fiducial cloud in the left-hand panel of Figure \ref{feedback:sfmass_photons}, this model is reasonably successful at reproducing the mass of Jeans Unstable gas measured in our simulations. It overestimates the mass loss for the $10^{47}$ s$^{-1}$ since the front stalls and is unable to ionise clumps in the direction of the densest parts of the cloud. We overpredict the Jeans Unstable mass in the $10^{49}$ s$^{-1}$ simulation after a few Myr. This is because the HII region expels some massive clumps from the simulation volume, meaning we are no longer able to track this mass. The more clumpy nature of the cloud without a magnetic field causes it to produce more Jeans unstable mass than the simulation with a magnetic field, as the cloud fragments more, while the ionisation front escapes preferentially in directions with lower densities.

For the More Compact and Most Compact clouds, our models overpredict the amount of mass evaporated. For the Most Compact cloud, the model with no photons (the upper light blue dotted line) fits better than the model with photons (the lower light blue dotted line). In the More Compact cloud, both models fail to reproduce the simulation results. As we find in the previous section, while the ionisation front in the More Compact cloud escapes the cloud, it does not affect the position of the dense clumps in the cloud. Thus our spherically-averaged model for the ionisation front radius overestimates the effect that the ionisation front has on the clumps inside the cloud.

Our measurements of the Jeans unstable mass in our simulations predict that the main effect of HII regions in clouds should be to evaporate star-forming clumps, reducing the star formation efficiency of the cloud. These measurements are backed up by a simple model based on \cite{Bertoldi1990}. However, in cases where the ionisation front stalls this simple model (which does not include the stalling radius) overpredicts the evaporation of dense clumps in the cloud. 

% \begin{figure*}
% \centerline{\includegraphics[width=0.48\hsize]{plots/photons/radiusmass.eps}
% \includegraphics[width=0.48\hsize]{plots/photons/massdistance.eps}}
% \caption{Properties of clumps in simulation \simname{N00\_B02} at 1.45 Myr (the point at which rapid accretion ends).}
%  \label{feedback:clumpproperties}
% \end{figure*}

\section{Discussion}
\label{discussion}

We now discuss the further consequences of our results and some limitations to our work. While our setup is deliberately simplified in various ways in order to make analytic comparisons possible, there are various aspects of feedback in molecular clouds that we omit that nonetheless are expected to play an important role.

One important effect that we leave out of our simulation is self-consistent star formation. Instead, we model emission from stars by assuming that a single black body source of photons is turned on in the centre of the cloud after one free-fall time. The most massive stars in a young cluster tend to be found in the centre of the cluster thanks to mass segregation \rev{\citep{Spitzer1969}}, and so it is not entirely inconsistent to use the approach given in this paper \rev{provided that the time that the most massive objects fall into the centre of the cloud is shorter than the time the ionisation front takes to escape the cloud. In practice, this criterion is best fit by the densest cloud, where the crossing time is shortest}.

However, in placing our source of photons by hand we miss three important points. The first is that we miss the early phase of star formation inside the clumps, in which the properties of the stars are sensitive to the radiative flux \citep[e.g.][]{Keto2002}. 

The second is that the emission rate of UV photons is roughly proportional to the cube of the mass of the most massive star in the cluster \citep{Vacca1996}, and thus we are unable to comment on the link between the cloud properties and the amount of feedback from UV photons in the cloud's star formation cycle. Observed ultracompact HII regions are expected to be short-lived \citep{Wood1989,Walsh1995}. Assuming observed clouds can reach the densities found in our compact cloud, one possibility is that the UV emission rate of the cluster formed in the cloud increases due to star formation until it is sufficient to break out of the cloud. This scenario requires simulations with self-consistent star formation to test, which we leave for future work.

The third is that our source is not gravitationally bound to the cloud, and this means that we overestimate the extent to which the stars are separated from the dense clumps in which they form. In addition to this, our cloud is modelled as an isolated sphere, and so star-forming clumps accrete only on the order of the free-fall time in the dense gas in the cloud. \cite{Dobbs2013} suggest that clouds should accrete as they pass through the spiral arms of galaxies.

In order to compensate for the lack of self-consistent star formation in our simulations, we locate regions of our simulations where there is Jeans unstable gas to predict where stars should form, and how much potentially star-forming mass should be lost by evaporation. We compare this to a simple model for  the evaporation of dense gas clumps based on \cite{Bertoldi1990}. When analysing the response of the Jeans unstable mass in our simulation to UV photoionisation, we do not find examples where the mass of unstable gas is increased by the presence of a UV source, and the feedback from UV photoionisation appears to be entirely negative (i.e. it reduces the mass available to form stars). We cannot discount a temporary boost in star formation rates from the initial compression of clouds as predicted by \cite{Walch2012a}, though. \cite{Dale2015} argues that we should be cautious in how we define ``triggered'' star formation. Similarly, we do not find that the compression of the cloud into a shell creates Jeans unstable gas, though this is still a subject of ongoing debate \citep[see, e.g. ][]{Kim2014}. \cite{Elmegreen1994,Hosokawa2006,Tremblin2014} predict that under certain regimes the shells around ionisation fronts should become unstable to fragmentation, though we do not find that this does not occur in our simulations. Even if the shell does not become unstable, the momentum transferred to the ISM from the expansion of the ionisation front can be anything up to that injected by a supernova, helping to maintain turbulence in the \ISM \citep{Gritschneder2009}. This in turn could help regulate star formation on a global \ISM scale over timescales longer than the life of the cloud, even if the influence of stellar feedback from UV photoionsation inside a cloud has a net negative effect.

The results of our model for the stalled expansion of ionisation fronts have some important consequences for star formation. Firstly, if ionisation fronts are unable to disrupt the cloud in which they form, they will be unable to prevent further star formation, at least until the point at which the emission rate from UV photons is sufficiently large to overcome the infall velocity. This is one possible explanation for the presence of observed ``super star clusters'' \citep[e.g.][]{Keto2005} - if massive, compact clouds are formed in starburst galaxies, photoionisation feedback will be unable to significantly disperse these clouds before they form the bulk of their stars. Secondly, the environment into which massive stars explode as supernovae changes the properties of the resulting blastwave. In circumstellar environments pre-processed by UV photoionisation, more energy is retained by the blastwave \citep[e.g.][]{Rogers2013,Geen2015,Walch2015}. \cite{Kimm2014} argue that if more momentum is deposited into a smaller portion of the \ISM \citep[see, e.g.,][]{Iffrig2015} then this can lead to a larger amount of energy from supernovae being transferred to the \ISM and galactic winds.

The presence of a magnetic field does not appear to strongly affect the bulk properties of the resulting HII region, even though it affects the structure of the cloud and HII region noticeably. If magnetic fields can alter the rate at which stars are formed by changing the structure of the densest parts of the cloud, then they would indeed alter the structure of HII regions by modifying the UV emission rate produced by the cluster embedded in the cloud.

There are various pre-supernova stellar feedback processes that occur in star-forming clouds other than UV photoionisation. Radiation pressure from infrared and reprocessed optical emission from stars can also aid cloud destruction, although there is still some debate as to the effectiveness of these processes \citep[e.g.][]{Krumholz2012c}. Radiation pressure has recently been added to \simname{RAMSES-RT} \citep{Rosdahl2015} and will be explored in future work. As well as radiative feedback, stars produce winds that can heat the gas around them. Some theoretical work has already been done by \cite{Garcia-Segura1996} in determining the effect these winds have on HII regions, particularly ultra-compact ones. For the majority of massive stars these winds are typically weak and provide a limited amount of energy \citep{Dale2014,Geen2015}, but more massive Wolf Rayet stars ($>$ 20-30 \Msolar) do produce significant hot bubbles in their circumstellar environment \citep{Dwarkadas2007}.

Another aspect not explored by these simulations is the role of metallicity, since we use only solar metallicity to calculate our cooling rates. Radiative cooling in primordial gas is particularly inefficient, and UV feedback at high redshift is expected to be even more efficient than in the simulations contained in this paper since the effects of metal cooling on the photoionised gas will be significantly reduced due to its low metallicity. Nonetheless, we expect the theoretical models given in this paper to extend to HII regions with different metallicities.

\section{Conclusions}
\label{conclusions}

\rev{We present a new set of analytic models that describe the evolution of HII regions in UV photoionisation equilibrium based on the arguments of \cite{SpitzerLyman1978,Dyson1980,Franco1990,Matzner2002,Raga2012}. We focus on cases where the cloud's density and velocity field corresponds to that expected in a virialised cloud supported by turbulence. We present limits in which the emergence of HII regions from their host cloud is prevented by pressure forces, either by thermal pressure or by ram pressure from gas flows. In order to determine the validity of these 1D models we compare them to a new suite of fully 3D simulations of UV photoionsation feedback in turbulent, magnetised, self-gravitating clouds. We perform these simulations using \simname{RAMSES-RT}, a Eulerian adaptive mesh refinement (AMR) radiation magnetohydrodynamics code.}

\rev{The presence of turbulence is important for two competing reasons. On one hand, it provides a ram pressure term that resists the expansion of HII regions in cases where the source of UV photons is weak or the cloud is dense enough. We calculate an analytic estimate for the radius at which turbulence prevents the HII region from expanding further. If this radius is smaller than the radius of the cloud, the HII region remains trapped and does not destroy the cloud. On the other hand, turbulence provides support for the cloud and prevents it from collapsing. If a cloud is dominated by radial flows from infalling gas, the density increases dramatically over time, requiring a much stronger source to destroy the cloud. Otherwise, the HII region collapses and is crushed by the cloud. Dense clumps orbiting the source position cause the HII region to ``flicker'' on and off as the dense gas efficiently absorbs the UV photons.}

\rev{The presence of a magnetic field limits fragmentation in the cloud, removing some channels of low-density gas through which the ionisation front can escape more easily, though the median radius of the ionisation front is similar in simulations with and without a magnetic field.}

We discuss briefly the expected role of UV photoionisation in regulating the star formation efficiency (SFE) of the cloud. Since we do not directly model star formation in this work, we instead measure the mass in Jeans unstable gas, which we find decreases with increasing UV emission rate. We compare this to a simple model for the evaporation of dense clumps by UV photons, and find a reasonable agreement in most cases, though the model overpredicts the mass evaporated for the weakest photon emission rates. We will address self-consistent star formation and the amount of UV radiation from these stars in a future work.

\section{Acknowlegements}
\label{acknowledgements}

We would like to thank the anonymous referee for their detailed and constructive comments, which improved upon the original manuscript and greatly helped clarify and strengthen the arguments made in this work. We also thank Romain Teyssier and Alex Richings for useful discussions during the preparation of this paper. The simulations presented here were run on the DiRAC facility jointly funded by STFC, the Large Facilities Capital Fund of BIS and the University of Oxford. We would like to warmly thank Jonathan Patterson for his help in running these simulations. This work has been funded by the the European Research Council under the European Community's Seventh Framework Programme (FP7/2007-2013). SG and PH are funded by Grant Agreement no. 306483 of this programme. PT is funded by Grant Agreement no. 247060, and JR by Grant Agreement 278594-GasAroundGalaxies. JR also acknowledges funding from the Marie Curie Training Network CosmoComp (PITN-GA-2009-238356).

% BIBLIOGRAPHY
 \bibliographystyle{mn2e}
 \bibliography{mcrtpaper}

\begin{thebibliography}{}

\bibitem[\protect\citeauthoryear{Alves, Lombardi \& Lada}{Alves
  et~al.}{2007}]{Alves2007}
Alves J.,  Lombardi M.,    Lada C.~J.,  2007, Astronomy and Astrophysics, 462,
  L17

\bibitem[\protect\citeauthoryear{Arthur, Henney, Mellema, {De Colle} \&
  V\'{a}zquez-Semadeni}{Arthur et~al.}{2011}]{Arthur2011}
Arthur S.~J.,  Henney W.~J.,  Mellema G.,  {De Colle} F.,
  V\'{a}zquez-Semadeni E.,  2011, Monthly Notices of the Royal Astronomical
  Society, 414, 1747

\bibitem[\protect\citeauthoryear{Audit \& Hennebelle}{Audit \&
  Hennebelle}{2005}]{Audit2005}
Audit E.,  Hennebelle P.,  2005, Astronomy and Astrophysics, 433, 1

\bibitem[\protect\citeauthoryear{Bacmann, Andre, Puget, Abergel, Bontemps \&
  Ward-Thompson}{Bacmann et~al.}{2000}]{Bacmann2000}
Bacmann A.,  Andre P.,  Puget J.-L.,  Abergel A.,  Bontemps S.,
  Ward-Thompson D.,  2000, Astronomy and Astrophysics, p.~28

\bibitem[\protect\citeauthoryear{Bertoldi \& McKee}{Bertoldi \&
  McKee}{1990}]{Bertoldi1990}
Bertoldi F.,  McKee C.~F.,  1990, The Astrophysical Journal, 354, 529

\bibitem[\protect\citeauthoryear{Bisbas, Haworth, Williams, Mackey, Tremblin,
  Raga, Arthur, Baczynski, Dale, Frostholm, Geen, Haugb\o~lle, Hubber, Iliev,
  Kuiper, Rosdahl, Sullivan, Walch \& W\"{u}nsch}{Bisbas
  et~al.}{2015}]{Bisbas2015}
Bisbas T.~G.,  Haworth T.~J.,  Williams R. J.~R.,  Mackey J.,  Tremblin P.,
  Raga a.~C.,  Arthur S.~J.,  Baczynski C.,  Dale J.~E.,  Frostholm T.,  Geen
  S.,  Haugb\o~lle T.,  Hubber D.,  Iliev I.~T.,  Kuiper R.,  Rosdahl J.,
  Sullivan D.,  Walch S.,    W\"{u}nsch R.,  2015, Monthly Notices of the Royal
  Astronomical Society, 453, 1324

\bibitem[\protect\citeauthoryear{Dale \& Bonnell}{Dale \&
  Bonnell}{2011}]{Dale2011}
Dale J.~E.,  Bonnell I.,  2011, Monthly Notices of the Royal Astronomical
  Society, 414, 321

\bibitem[\protect\citeauthoryear{Dale, Bonnell, Clarke \& Bate}{Dale
  et~al.}{2005}]{Dale2005}
Dale J.~E.,  Bonnell I.~A.,  Clarke C.~J.,    Bate M.~R.,  2005, Monthly
  Notices of the Royal Astronomical Society, 358, 291

\bibitem[\protect\citeauthoryear{Dale, Ercolano \& Bonnell}{Dale
  et~al.}{2012}]{Dale2012}
Dale J.~E.,  Ercolano B.,    Bonnell I.~A.,  2012, Monthly Notices of the Royal
  Astronomical Society, 424, 377

\bibitem[\protect\citeauthoryear{Dale, Ercolano \& Bonnell}{Dale
  et~al.}{2013}]{Dale2013}
Dale J.~E.,  Ercolano B.,    Bonnell I.~A.,  2013, Monthly Notices of the Royal
  Astronomical Society, 430, 234

\bibitem[\protect\citeauthoryear{Dale, Haworth \& Bressert}{Dale
  et~al.}{2015}]{Dale2015}
Dale J.~E.,  Haworth T.~J.,    Bressert E.,  2015, Monthly Notices of the Royal
  Astronomical Society, 450, 1199

\bibitem[\protect\citeauthoryear{Dale, Ngoumou, Ercolano \& Bonnell}{Dale
  et~al.}{2014}]{Dale2014}
Dale J.~E.,  Ngoumou J.,  Ercolano B.,    Bonnell I.~A.,  2014, Monthly Notices
  of the Royal Astronomical Society, 442, 694

\bibitem[\protect\citeauthoryear{Didelon, Motte, Tremblin \& Hill}{Didelon
  et~al.}{2015}]{Didelon2015}
Didelon P.,  Motte F.,  Tremblin P.,    Hill T.,  2015, Astronomy and
  Astrophysics, accepted

\bibitem[\protect\citeauthoryear{Dobbs \& Pringle}{Dobbs \&
  Pringle}{2013}]{Dobbs2013}
Dobbs C.~L.,  Pringle J.~E.,  2013, Monthly Notices of the Royal Astronomical
  Society, 432, 653

\bibitem[\protect\citeauthoryear{Draine}{Draine}{2011}]{Draine2011}
Draine Â.,  2011, Physics of the Interstellar and Intergalactic Medium by Bruce
  T. Draine. Princeton University Press

\bibitem[\protect\citeauthoryear{Dwarkadas}{Dwarkadas}{2007}]{Dwarkadas2007}
Dwarkadas V.~V.,  2007, The Astrophysical Journal, 667, 226

\bibitem[\protect\citeauthoryear{Dyson \& Williams}{Dyson \&
  Williams}{1980}]{Dyson1980}
Dyson Â.,  Williams Â.,  1980, New York

\bibitem[\protect\citeauthoryear{Elmegreen}{Elmegreen}{1994}]{Elmegreen1994}
Elmegreen B.~G.,  1994, The Astrophysical Journal, 427, 384

\bibitem[\protect\citeauthoryear{Franco, Tenorio-Tagle \& Bodenheimer}{Franco
  et~al.}{1990}]{Franco1990}
Franco J.,  Tenorio-Tagle G.,    Bodenheimer P.,  1990, The Astrophysical
  Journal, 349, 126

\bibitem[\protect\citeauthoryear{Fromang, Hennebelle \& Teyssier}{Fromang
  et~al.}{2006}]{Fromang2006}
Fromang S.,  Hennebelle P.,    Teyssier R.,  2006, Astronomy and Astrophysics,
  457, 371

\bibitem[\protect\citeauthoryear{Gao, Xu \& Law}{Gao et~al.}{2015}]{Gao2015}
Gao Y.,  Xu H.,    Law C.~K.,  2015, The Astrophysical Journal, 799, 227

\bibitem[\protect\citeauthoryear{Garcia-Segura \& Franco}{Garcia-Segura \&
  Franco}{1996}]{Garcia-Segura1996}
Garcia-Segura G.,  Franco J.,  1996, The Astrophysical Journal, 469, 171

\bibitem[\protect\citeauthoryear{Geen, Rosdahl, Blaizot, Devriendt \&
  Slyz}{Geen et~al.}{2015}]{Geen2015}
Geen S.,  Rosdahl J.,  Blaizot J.,  Devriendt J.,    Slyz A.,  2015, Monthly
  Notices of the Royal Astronomical Society, 448, 3248

\bibitem[\protect\citeauthoryear{Gritschneder, Naab, Walch, Burkert \&
  Heitsch}{Gritschneder et~al.}{2009}]{Gritschneder2009}
Gritschneder M.,  Naab T.,  Walch S.,  Burkert A.,    Heitsch F.,  2009, The
  Astrophysical Journal, 694, L26

\bibitem[\protect\citeauthoryear{Hennebelle}{Hennebelle}{2013}]{Hennebelle2013}
Hennebelle P.,  2013, Astronomy \& Astrophysics, 556, A153

\bibitem[\protect\citeauthoryear{Hosokawa \& Inutsuka}{Hosokawa \&
  Inutsuka}{2006}]{Hosokawa2006}
Hosokawa T.,  Inutsuka S.,  2006, The Astrophysical Journal, 646, 240

\bibitem[\protect\citeauthoryear{Hunter}{Hunter}{1962}]{Hunter1962}
Hunter C.,  1962, The Astrophysical Journal, 136, 594

\bibitem[\protect\citeauthoryear{Iffrig \& Hennebelle}{Iffrig \&
  Hennebelle}{2015}]{Iffrig2015}
Iffrig O.,  Hennebelle P.,  2015, Astronomy \& Astrophysics, 576, A95

\bibitem[\protect\citeauthoryear{Johnstone, Hollenbach \& Bally}{Johnstone
  et~al.}{1998}]{Johnstone1998}
Johnstone Â.,  Hollenbach Â.,    Bally Â.,  1998, The Astrophysical Journal,
  499, 758

\bibitem[\protect\citeauthoryear{Joung \& {Mac Low}}{Joung \& {Mac
  Low}}{2006}]{Joung2006}
Joung M. K.~R.,  {Mac Low} M.,  2006, The Astrophysical Journal, 653, 1266

\bibitem[\protect\citeauthoryear{Kahn}{Kahn}{1954}]{KahnF.D.1954}
Kahn F.~D.,  1954, Bulletin of the Astronomical Institutes of the Netherlands,
  12

\bibitem[\protect\citeauthoryear{Keto}{Keto}{2002}]{Keto2002}
Keto E.,  2002, The Astrophysical Journal, 580, 980

\bibitem[\protect\citeauthoryear{Keto}{Keto}{2003}]{Keto2003}
Keto E.,  2003, The Astrophysical Journal, 599, 1196

\bibitem[\protect\citeauthoryear{Keto, Ho \& Lo}{Keto et~al.}{2005}]{Keto2005}
Keto E.,  Ho L.~C.,    Lo K.,  2005, The Astrophysical Journal, 635, 1062

\bibitem[\protect\citeauthoryear{Kim \& Kim}{Kim \& Kim}{2014}]{Kim2014}
Kim J.-G.,  Kim W.-T.,  2014, The Astrophysical Journal, 797, 135

\bibitem[\protect\citeauthoryear{Kimm \& Cen}{Kimm \& Cen}{2014}]{Kimm2014}
Kimm T.,  Cen R.,  2014, The Astrophysical Journal, 788, 121

\bibitem[\protect\citeauthoryear{Krumholz, Stone \& Gardiner}{Krumholz
  et~al.}{2007}]{Krumholz2007a}
Krumholz M.~R.,  Stone J.~M.,    Gardiner T.~A.,  2007, The Astrophysical
  Journal, 671, 518

\bibitem[\protect\citeauthoryear{Krumholz \& Thompson}{Krumholz \&
  Thompson}{2012}]{Krumholz2012c}
Krumholz M.~R.,  Thompson T.~A.,  2012, The Astrophysical Journal, 760, 155

\bibitem[\protect\citeauthoryear{Larson}{Larson}{1969}]{Larson1969}
Larson Â.,  1969, Monthly Notices of the Royal Astronomical Society, 145

\bibitem[\protect\citeauthoryear{Lefloch \& Lazareff}{Lefloch \&
  Lazareff}{1994}]{Lefloch1994}
Lefloch Â.,  Lazareff Â.,  1994, Astronomy and Astrophysics (ISSN 0004-6361),
  289, 559

\bibitem[\protect\citeauthoryear{Martins, Schaerer \& Hillier}{Martins
  et~al.}{2005}]{Martins2005}
Martins F.,  Schaerer D.,    Hillier D.~J.,  2005, Astronomy and Astrophysics,
  436, 1049

\bibitem[\protect\citeauthoryear{Matzner}{Matzner}{2002}]{Matzner2002}
Matzner C.~D.,  2002, The Astrophysical Journal, 566, 302

\bibitem[\protect\citeauthoryear{Matzner \& McKee}{Matzner \&
  McKee}{2000}]{Matzner2000}
Matzner C.~D.,  McKee C.~F.,  2000, The Astrophysical Journal, 545, 364

\bibitem[\protect\citeauthoryear{Mellema, Arthur, Henney, Iliev \&
  Shapiro}{Mellema et~al.}{2006}]{Mellema2006}
Mellema G.,  Arthur S.~J.,  Henney W.~J.,  Iliev I.~T.,    Shapiro P.~R.,
  2006, The Astrophysical Journal, 647, 397

\bibitem[\protect\citeauthoryear{Nielbock, Launhardt, Steinacker, Stutz, Balog,
  Beuther, Bouwman, Henning, Hily-Blant, Kainulainen, Krause, Linz, Lippok,
  Ragan, Risacher \& Schmiedeke}{Nielbock et~al.}{2012}]{Nielbock2012}
Nielbock M.,  Launhardt R.,  Steinacker J.,  Stutz a.~M.,  Balog Z.,  Beuther
  H.,  Bouwman J.,  Henning T.,  Hily-Blant P.,  Kainulainen J.,  Krause O.,
  Linz H.,  Lippok N.,  Ragan S.,  Risacher C.,    Schmiedeke a.,  2012,
  Astronomy \& Astrophysics, 547, A11

\bibitem[\protect\citeauthoryear{Oort \& Spitzer}{Oort \&
  Spitzer}{1955}]{Oort1955}
Oort J.~H.,  Spitzer L.,  1955, The Astrophysical Journal, 121, 6

\bibitem[\protect\citeauthoryear{Pecaut, Mamajek \& Bubar}{Pecaut
  et~al.}{2012}]{Pecaut2012}
Pecaut M.~J.,  Mamajek E.~E.,    Bubar E.~J.,  2012, The Astrophysical Journal,
  746, 154

\bibitem[\protect\citeauthoryear{Peretto, Fuller, Duarte-Cabral, Avison,
  Hennebelle, Pineda, Andr\'{e}, Bontemps, Motte, Schneider \&
  Molinari}{Peretto et~al.}{2013}]{Peretto2013}
Peretto N.,  Fuller G.~A.,  Duarte-Cabral A.,  Avison A.,  Hennebelle P.,
  Pineda J.~E.,  Andr\'{e} P.,  Bontemps S.,  Motte F.,  Schneider N.,
  Molinari S.,  2013, Astronomy \& Astrophysics, 555, A112

\bibitem[\protect\citeauthoryear{Raga, Canto \& Rodriguez}{Raga
  et~al.}{2012}]{Raga2012}
Raga A.~C.,  Canto J.,    Rodriguez L.~F.,  2012, Monthly Notices of the Royal
  Astronomical Society, 419, L39

\bibitem[\protect\citeauthoryear{Rogers \& Pittard}{Rogers \&
  Pittard}{2013}]{Rogers2013}
Rogers H.,  Pittard J.~M.,  2013, Monthly Notices of the Royal Astronomical
  Society, 431, 1337

\bibitem[\protect\citeauthoryear{Rosdahl, Blaizot, Aubert, Stranex \&
  Teyssier}{Rosdahl et~al.}{2013}]{Rosdahl2013}
Rosdahl J.,  Blaizot J.,  Aubert D.,  Stranex T.,    Teyssier R.,  2013,
  Monthly Notices of the Royal Astronomical Society, 436, 2188

\bibitem[\protect\citeauthoryear{Rosdahl \& Teyssier}{Rosdahl \&
  Teyssier}{2015}]{Rosdahl2015}
Rosdahl J.,  Teyssier R.,  2015, Monthly Notices of the Royal Astronomical
  Society, 449, 4380

\bibitem[\protect\citeauthoryear{Saff \& Kuijlaars}{Saff \&
  Kuijlaars}{1997}]{Saff1997}
Saff E.~B.,  Kuijlaars A. B.~J.,  1997, The Mathematical Intelligencer, 19, 5

\bibitem[\protect\citeauthoryear{Schaller, Schaerer, Meynet \& Maeder}{Schaller
  et~al.}{1992}]{Schaller1992}
Schaller Â.,  Schaerer Â.,  Meynet Â.,    Maeder Â.,  1992, Astronomy and
  Astrophysics Supplement Series (ISSN 0365-0138), 96, 269

\bibitem[\protect\citeauthoryear{Soler, Hennebelle, Martin,
  Miville-Desch\^{e}nes, Netterfield \& Fissel}{Soler et~al.}{2013}]{Soler2013}
Soler J.~D.,  Hennebelle P.,  Martin P.~G.,  Miville-Desch\^{e}nes M.-A.,
  Netterfield C.~B.,    Fissel L.~M.,  2013, The Astrophysical Journal, 774,
  128

\bibitem[\protect\citeauthoryear{Spitzer Lyman}{Spitzer}{1969}]{Spitzer1969}
Spitzer Lyman J.,  1969, The Astrophysical Journal, 158, L139

\bibitem[\protect\citeauthoryear{Spitzer}{Spitzer}{1978}]{SpitzerLyman1978}
Spitzer L.,  1978, Physical Processes in the Interstellar Medium, New York
  Wiley-Interscience

\bibitem[\protect\citeauthoryear{Sternberg, Hoffmann \& Pauldrach}{Sternberg
  et~al.}{2003}]{Sternberg2003}
Sternberg A.,  Hoffmann T.~L.,    Pauldrach A. W.~A.,  2003, The Astrophysical
  Journal, 599, 1333

\bibitem[\protect\citeauthoryear{Sutherland \& Dopita}{Sutherland \&
  Dopita}{1993}]{Sutherland1993}
Sutherland R.~S.,  Dopita M.~A.,  1993, The Astrophysical Journal Supplement
  Series, 88, 253

\bibitem[\protect\citeauthoryear{Teyssier}{Teyssier}{2002}]{Teyssier:2002p533}
Teyssier R.,  2002, Astronomy and Astrophysics, 385, 337

\bibitem[\protect\citeauthoryear{Tremblin, Anderson, Didelon, Raga, Minier,
  Ntormousi, Pettitt, Pinto, Samal, Schneider \& Zavagno}{Tremblin
  et~al.}{2014}]{Tremblin2014a}
Tremblin P.,  Anderson L.~D.,  Didelon P.,  Raga a.~C.,  Minier V.,  Ntormousi
  E.,  Pettitt a.,  Pinto C.,  Samal M.~R.,  Schneider N.,    Zavagno a.,
  2014, Astronomy \& Astrophysics, 568, A4

\bibitem[\protect\citeauthoryear{Tremblin, Audit, Minier, Schmidt \&
  Schneider}{Tremblin et~al.}{2012}]{Tremblin2012}
Tremblin P.,  Audit E.,  Minier V.,  Schmidt W.,    Schneider N.,  2012,
  Astronomy \& Astrophysics, 546, A33

\bibitem[\protect\citeauthoryear{Tremblin, Schneider, Minier, Didelon, Hill \&
  Others}{Tremblin et~al.}{2014}]{Tremblin2014}
Tremblin P.,  Schneider N.,  Minier V.,  Didelon P.,  Hill T.,    Others A.,
  2014, Astronomy \& Astrophysics, 564, A106

\bibitem[\protect\citeauthoryear{Vacca, Garmany \& Shull}{Vacca
  et~al.}{1996}]{Vacca1996}
Vacca W.~D.,  Garmany C.~D.,    Shull J.~M.,  1996, The Astrophysical Journal,
  460, 914

\bibitem[\protect\citeauthoryear{Walch \& Naab}{Walch \&
  Naab}{2015}]{Walch2015}
Walch S.,  Naab T.,  2015, Monthly Notices of the Royal Astronomical Society,
  451, 2757

\bibitem[\protect\citeauthoryear{Walch, Whitworth, Bisbas, Wunsch \&
  Hubber}{Walch et~al.}{2013}]{Walch2013}
Walch S.,  Whitworth A.~P.,  Bisbas T.~G.,  Wunsch R.,    Hubber D.~A.,  2013,
  Monthly Notices of the Royal Astronomical Society, -1, 2099

\bibitem[\protect\citeauthoryear{Walch, Whitworth \& Girichidis}{Walch
  et~al.}{2012}]{Walch2012a}
Walch S.,  Whitworth A.~P.,    Girichidis P.,  2012, Monthly Notices of the
  Royal Astronomical Society, 419, 760

\bibitem[\protect\citeauthoryear{Walch, Whitworth, Bisbas, W\"{u}nsch \&
  Hubber}{Walch et~al.}{2012}]{Walch2012}
Walch S.~K.,  Whitworth A.~P.,  Bisbas T.,  W\"{u}nsch R.,    Hubber D.,  2012,
  Monthly Notices of the Royal Astronomical Society, 427, 625

\bibitem[\protect\citeauthoryear{Walsh, Lyland, Robinson, Bourke \&
  James}{Walsh et~al.}{1995}]{Walsh1995}
Walsh A.~J.,  Lyland A.~R.,  Robinson G.,  Bourke T.~L.,    James S.~D.,  1995,
  Publications of the Astronomical Society of Australia, 12, 186

\bibitem[\protect\citeauthoryear{Ward-Thompson, Scott, Hills \&
  Andre}{Ward-Thompson et~al.}{1994}]{Ward-Thompson1994}
Ward-Thompson Â.,  Scott Â.,  Hills Â.,    Andre Â.,  1994, Monthly Notices of
  the Royal Astronomical Society, 268

\bibitem[\protect\citeauthoryear{Whitworth \& Zinnecker}{Whitworth \&
  Zinnecker}{2004}]{Whitworth2004}
Whitworth A.~P.,  Zinnecker H.,  2004, Astronomy and Astrophysics, 427, 299

\bibitem[\protect\citeauthoryear{Whitworth}{Whitworth}{1979}]{Whitworth1979}
Whitworth Â.,  1979, Monthly Notices of the Royal Astronomical Society, 186, 59

\bibitem[\protect\citeauthoryear{Williams \& McKee}{Williams \&
  McKee}{1997}]{Williams1997}
Williams Â.,  McKee Â.,  1997, The Astrophysical Journal, 476, 166

\bibitem[\protect\citeauthoryear{Wood \& Churchwell}{Wood \&
  Churchwell}{1989}]{Wood1989}
Wood D. O.~S.,  Churchwell E.,  1989, The Astrophysical Journal Supplement
  Series, 69, 831

\bibitem[\protect\citeauthoryear{Yun \& Clemens}{Yun \&
  Clemens}{1991}]{Yun1991}
Yun J.~L.,  Clemens D.~P.,  1991, The Astrophysical Journal, 381, 474

\end{thebibliography}

\appendix

\section{Front Expansion in a Turbulent Medium Derivation}
\label{infalllimited}

In this analysis we allow the velocity of the ionisation front and the shock to differ (in Section \ref{expansion:powerlaw} we assume they are the same). We write the Rankine Hugoniot conditions (describing the conditions across a shock interface) in the frame of the shock below, where $v_s$ is the shock velocity, and $n_c$ and $v_c$ are the post-shock number density and velocity (all other symbols as defined in Section \ref{expansion}:

\begin{equation} 
n_{ext}(r,t)(v_s-v_{ext}(r,t)) = n_c v_c 
\end{equation}
\begin{equation}
 n_{ext}(r,t)((v_s-v_{ext}(r,t))^2+c_{ext}^2) = n_c (v_c^2+c_{ext}^2) 
 \end{equation}
which can be combined to get: 
\begin{equation} M_{ext}^2 \equiv \frac{n_c}{n_{ext}} = \left(\frac{v_s-v_{ext}(r,t)}{c_{ext}}\right)^2 \end{equation}
hence, 
\begin{equation} 
v_c = \frac{c_{ext}^2}{v_s-v_{ext}(r,t)}. 
\end{equation} Then from Raga et al. (2012), we can write the relation between the velocity of the ionisation front and the velocity of the shock:

\begin{equation} 
v_s = \frac{\mathrm{d}r_i}{\mathrm{d}t}+v_c
\end{equation}
and so \begin{equation}
v_s = \frac{\mathrm{d}r_i}{\mathrm{d}t}+\frac{c_{ext}^2}{v_s-v_{ext}(r,t)} 
\end{equation}
giving \begin{equation}
 \frac{\mathrm{d}r_i}{\mathrm{d}t} = (v_s-v_{ext}(r,t))-\frac{c_{ext}^2}{v_s-v_{ext}(r,t)}+v_{ext}(r,t). 
 \end{equation}
 
We now find an expression for $(v_s-v_{ext}(r,t))$. As in Equation \ref{expansion:photon_balance}, photoionisation equilibrium gives 
\begin{equation}
n_i^2r_i^3=3\int_0^{r_s} n_{ext}(r,0)^2 r^2\mathrm{d}r=n_0^2r_s^3.
\label{infalllimited:photobalance}
\end{equation}

Momentum conservation at the ionisation front gives:
\begin{equation}
 n_ic_i^2 = n_{ext}(r,t)(v_s-v_{ext}(r,t))^2
\end{equation}

Using Equation \ref{infalllimited:photobalance}, this becomes:
\begin{equation}
\left(\frac{r_{s}}{r_i}\right)^{3/2} = \frac{n_{ext}(r,t)}{n_0}\frac{(v_s-v_{ext}(r,t))^2}{c_i^2}
\end{equation}
and so
\begin{equation}
\frac{v_s-v_{ext}(r,t)}{c_i} = \left(\frac{r_{s}}{r_i}\right)^{3/4}\left(\frac{n_0}{n_{ext}(r,t)}\right)^{1/2}
\end{equation}

Hence the equation for the velocity of the ionisation front is:

\begin{equation}
\frac{1}{c_i}\frac{\mathrm{d}r_i}{\mathrm{d}t} = F(r,t)-\frac{c_{ext}^2}{c_i^2}\frac{1}{F(r,t)}+\frac{v_{ext}(r,t)}{c_i} ,
\end{equation} where
\begin{equation}
F(r,t) \equiv \left(\frac{r_{s}}{r_i}\right)^{3/4}\left(\frac{n_0}{n_{ext}(r,t)}\right)^{1/2}
 \end{equation}

\section{Cloud Outflow Model}
\label{cloudoutflow}

We now discuss a model that takes into account the difference in distance from the source to the cloud edge with angle, $r_{cloud}$. We sample $r_{cloud}$ using rays along the vectors given in Appendix \ref{appendix:lossampling}. We find that $r_{cloud}$ has a roughly uniform probability distribution function with limits at 3 and 12 pc in the Fiducial cloud.

We solve the radial evolution of the ionisation front up to $r_{cloud}$ using Equation \ref{overview:radius}, with $w=0$, i.e. a flat density field (since the spherically-averaged power law model no longer applies) with $n_0$ set to the mean density of the cloud. We set $t_{cloud}$ to the time that the solution reaches $r_{cloud}$. At this point, the ionisation front enters the diffuse external medium. We add a step function to this density profile where for $r > r_{cloud}$, the density of the external medium is given by $n_d = 1~$atom cm$^{-3}$.

We modify Equation \ref{expansion:ram_pressure} to give
\begin{equation}
  n_{d} \dot{r_i}^2 = n_i c_i^2.
  \label{cloudoutflow:ram_pressure}
\end{equation} We do not change Equation \ref{expansion:photon_balance} since photoionisation equilibrium still holds, and none of these terms depends on the density of the external medium after $t=0$. Solving as in Section \ref{expansion:powerlaw}, we find
\begin{equation}
\begin{split}
 r_i (t > t_{cloud}) = \\
r_{cloud} \left(1 + \frac{7}{4}
\left( \frac{n_0}{n_d}\right )^{1/2}
\left(\frac{r_s}{r_{cloud}}\right )^{3/4}
\frac{c_i (t - t_{cloud})}{r_{cloud}} \right)^{4/7}.
\end{split}
\label{cloudoutflow:expansion}
\end{equation}

\begin{figure}
 \centerline{\includegraphics[width=0.98\hsize]{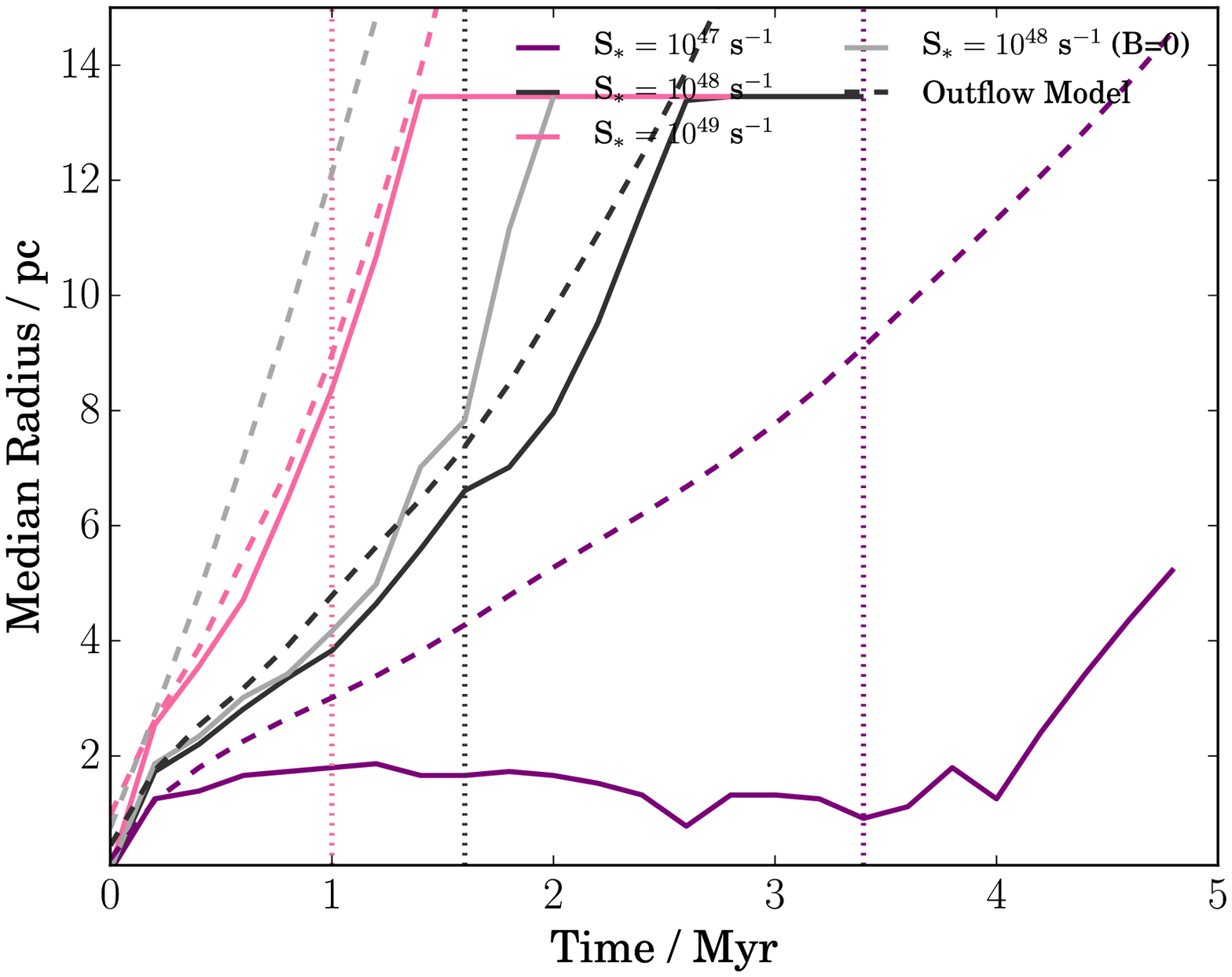}}
 \centerline{\includegraphics[width=0.98\hsize]{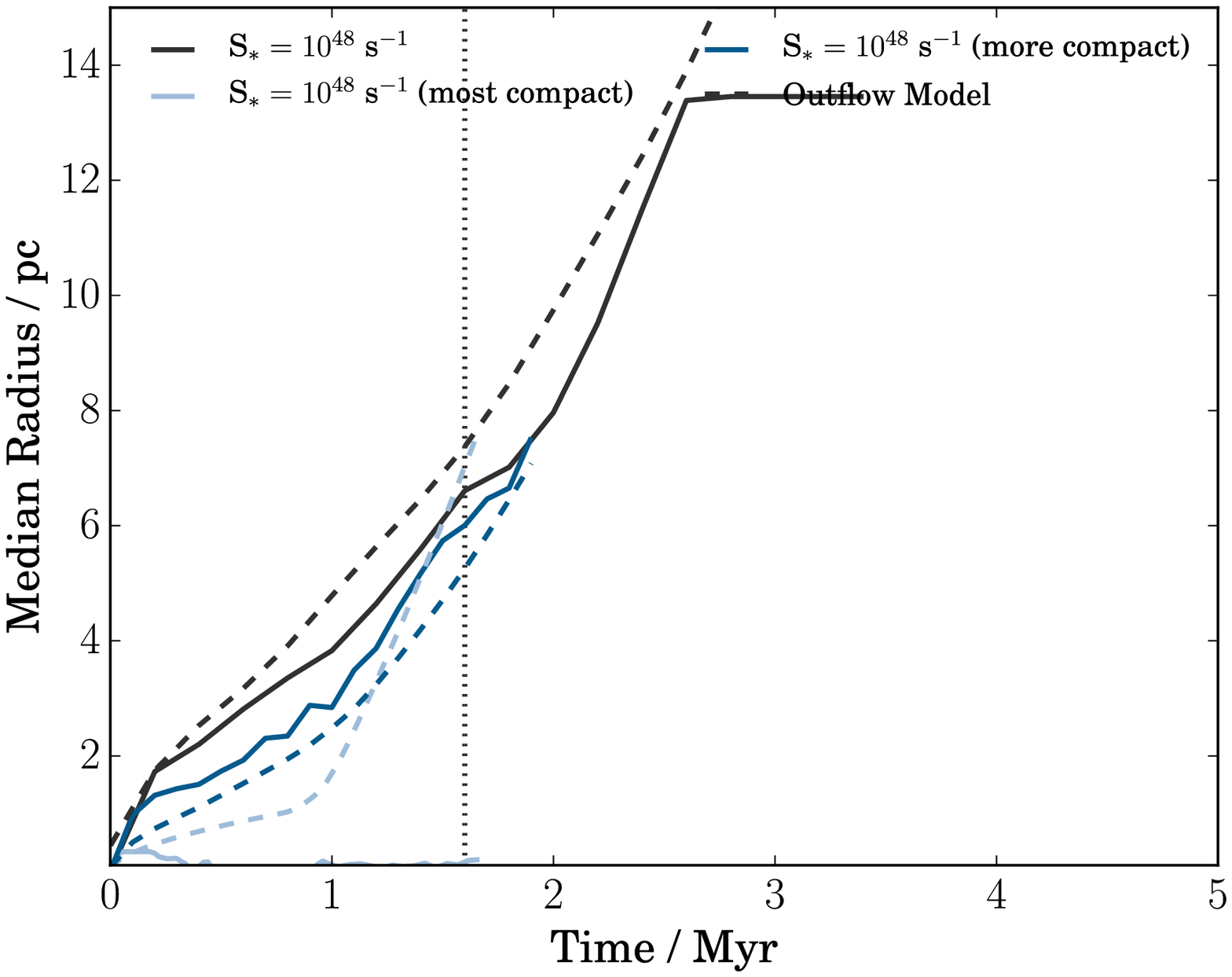}}
 \caption{Comparison of the ``Outflow'' model with the mean radius found in the simulation results as in Figure \protect\ref{flux:properties}. Upper panel: varying photon flux. Lower panel: varying cloud compactness.} 
 \label{cloudoutflow:mediandradius}
 \end{figure}

We solve this for $r_{cloud}$ varying from 3 to 12 pc, giving a distribution of solutions for $r_i$. The median value has $r_{cloud}$ = 7.5 pc. We plot these solutions for each of the simulations in the Fiducial cloud in Figure \ref{cloudoutflow:mediandradius}.

The fit to this model is better than that of the power law for \simname{N48\_B02} and \simname{N49\_B02}, which smooths over the sharp discontinuity at $r_c$ in cloud density. However, the model fails in the case of \simname{N48\_B00}, which has a density profile that is poorly fit by a flat field inside the cloud, where the ``Power Law'' model in Section \ref{expansion:powerlaw} is a better fit. The model also ignores clump motions and thus does not match \simname{N47\_B02} or \simname{N48\_B02\_C} any better than the ``Power Law'' model. In addition, once the ionisation front has left the cloud its expansion no longer affects significantly the properties inside the cloud - for example, the momentum and ionised mass are well captured by the power law model. However, in cases where the expansion of the front once it leaves the cloud is important, this two-step outflow model is a useful tool in understanding the HII region.

\section{Shell Instability}
\label{feedback:shell}

In Section \ref{results:feedback} we describe the loss of potentially star-forming mass from dense clumps as they are evaporated by the UV photons. However, as explained in Sections \ref{expansion:powerlaw} and \ref{results:free:ionmassandmom}, the amount of mass in ionised gas is typically much smaller than the mass of neutral gas displaced by the ionisation front. Most of the mass displaced thus ends up in the dense, neutral shell around the ionisation front. Here we discuss the possibility that this shell becomes Jeans Unstable and thus able to fragment and form stars.

\cite{Elmegreen1994} states that a dense shell around a shock becomes unstable to fragmentation when 
\begin{equation}
\frac{\pi G \rho_0}{3 c_0} > \frac{8^{1/2} V_{shell}}{r^2_{shell}}
\label{shell_stability}
\end{equation} where $\rho_0$ is the initial density, $c_0$ is the sound speed in the shell, and $V_{shell}$ and $r_{shell}$ are the speed and radius of the shell respectively. We can assume these are $\dot{r_i}$ and $r_i$ as before. Differentiating Equation \ref{overview:radius}), we get $\dot{r_i} = \psi r_i / t$. For an initial density $\rho_0 = m_H / X \times 500~$\atcc with $c_0=0.4$ km/s, we find that this criterion becomes
\begin{equation}
r_{shell} t > 100~\mathrm{pc Myr}
\label{shell_stability_ours}
\end{equation} for the Fiducial cloud. This requires the shell to travel at least 25 pc in 4 Myr (roughly the lifetime of the most massive star in our simulation) before it becomes unstable to fragmentation. We thus do not expect the shells around our ionisation fronts to become unstable in our simulations. However, on longer timescales it is possible that the shells would fragment and form stars. Another possibility is that this shell would encounter another molecular cloud, triggering star formation via shock compression of this cloud.

% HACK - force link to be on whole page
\newpage

\section{Line-of-Sight Sampling}
\label{appendix:lossampling}

We sample lines of sight on a sphere according to the following algorithm \citep{Saff1997}\footnote{see also \url{http://people.sc.fsu.edu/~jburkardt/f_src/sphere_grid/sphere_grid.html}}. For $N$ lines of sight with index $i=\{1,...,N\}$, the $i$th line of sight is defined as:
\begin{equation}
 \phi_i = \arccos \left( \frac{2 i - N - 1}{N-1} \right)
\end{equation}
\begin{equation}
\theta_i = \sum_{n=2}^{i} \left( \frac{3.6}{\sin\phi_n\sqrt{N}} \right)
\end{equation} where $\theta_1 = \theta_n = 0$.
The points on the surface of the sphere of unit radius $\mathbf{r}_i$ are then defined as
\begin{equation}
 \mathbf{r}_i =
\begin{bmatrix}
\sin \phi_i \cos \theta_i \\
\sin \phi_i \sin \theta_i \\
\cos \phi_i
\end{bmatrix}
\end{equation}

\end{document}